\documentclass[12pt]{article}
\usepackage{bbm}
\usepackage{pifont}
\usepackage{mathrsfs}
\usepackage{amsfonts}
\usepackage{graphicx}
\usepackage{subfigure}
\usepackage{caption}
\usepackage{latexsym}
\usepackage{epsfig}
\usepackage{amsmath}
\usepackage{amssymb}
\usepackage{color}
\usepackage{amsthm}

\newtheorem{theo}{\sc Theorem}
\newtheorem{lemm}{\sc Lemma}[section]
\newtheorem{rema}{\sc Remark}

\makeatletter
\@addtoreset{equation}{section}
\makeatother

\renewcommand{\baselinestretch}{1.2}
\title{Unsupervised Learning of Mixture Regression Models for Longitudinal Data}

\author{\sc  Peirong Xu$^{1}$, Heng Peng$^2$, Tao Huang$^3$\footnote{{\bf Corresponding Author:}
Tao Huang, Associate Professor, School of Statistics and Management,
      Shanghai University of Finance and Economics, Shanghai, China, 200433.
(\textit{E-mail:huang.tao@mail.shufe.edu.cn}). }
              \\ {\footnotesize$^{1}$ College of Mathematics and Sciences, Shanghai Normal University, Shanghai, China}
              \\        {\footnotesize $ ^{2}$ Department of Mathematics,
      Hong Kong Baptist University, Hong Kong, China}
              \\        {\footnotesize $ ^{3}$ School of Statistics and Management,
      Shanghai University of Finance and Economics, Shanghai, China}
       \\}

\date{}

\begin{document}
\maketitle

\setlength{\baselineskip}{24pt}
\renewcommand{\baselinestretch}{1.2}

\begin{quotation}
\noindent { \bf Abstract:}
This paper is concerned with learning of mixture regression models for individuals that are measured repeatedly. The adjective ``unsupervised" implies that the number of mixing components is unknown and has to be determined, ideally by data driven tools. For this purpose, a novel penalized method is proposed to simultaneously select the number of mixing components and to estimate the mixture proportions and unknown parameters in the models. The proposed method is capable of handling both continuous and discrete responses by only requiring the first two moment conditions of the model distribution. It is shown to be consistent in both selecting the number of components and estimating the mixture proportions and unknown regression parameters. Further, a modified EM algorithm is developed to seamlessly integrate model selection and estimation. Simulation studies are conducted to evaluate the finite sample performance of the proposed procedure. And it is further illustrated via an analysis of a primary biliary cirrhosis data set.
\vspace{9pt}


\par
\end{quotation}\par

\section{Introduction}\label{sec1}

In many medical studies, the marker of disease progression and a variety of characteristics are routinely measured during the patients' follow-up visit to decide on future treatment actions. Consider a motivating Mayo Clinic trial with primary biliary cirrhosis (PBC), wherein a number of serological, clinical and histological parameters were recorded for each of 312 patients from 1974 to 1984. This longitudinal study had a median follow-up time of 6.3 years as some patients missed their appointments due to worsening medical condition of some labs. It is known that PBC is a fatal chronic cholesteric liver disease, which is characterized histopathologically by portal inflammation and immune-mediated destruction of the intrahepatic bile ducts (Pontecorvo, Levinson, and Roth, 1992). It can be divided into four histologic stages, but with nonuniformly affected liver. The diagnosis of PBC is important for the medical treatment with Ursodiol has been shown to halt disease progression and improve survival without need for liver transplantation (Talwalkar and Lindor, 2003). Therefore, one goal of the study was the investigation of the serum bilirubin level, an important marker of PBC progression, in relation to the time and to potential clinical and histological covariates. Another issue that should be accounted for is the unobservable heterogeneity between subjects that may not be explained by the covariates. The changes in inflammation and bile ducts occur at different rates and with varying degrees of severity in different patients, so the heterogeneous patients could potentially belong to different latent groups. To address these problems, there is a demand for mixture regression modeling for subjects on the basis of longitudinal measurements.

There are various research works on mixture regression models for longitudinal outcome data, particularly in the context of model-based probabilistic clustering (Fraely and Raftery, 2002). For example, De la Cruz-Mes$\acute{i}$a et. al. (2008) proposed a mixture of non-linear hierarchical models with Gaussian subdistributions;
McNicholas and Murphy (2010) extended the Gaussian mixture models with Cholesky-decomposed group covariance structure;
Kom$\acute{a}$rek and Kom$\acute{a}$rkov$\acute{a}$ (2013) introduced a generalized linear mixed model for components'
densities under the Gaussian mixture framework; Heinzl and Tutz (2013) considered linear mixed models with approximate
Dirichlet process mixtures. Other relevant work includes Celeux et. al. (2005), Booth et. al. (2008), Pickles and Croudace
(2010), Maroutti (2011), Erosheva et. al. (2014) and some of the references therein. Compared with heuristic methods such as
the k-means method (Genolini and Falissard, 2010), issues like the selection of the number of clusters (or components) can
be addressed in a principled way. However, most of them assume a parametric mixture distribution, which may be too
restrictive and invalid in practice when the true data-generating mechanism indicates otherwise.

A key concern for the performance of mixture modeling is the selection of the number of components. A mixture with too many
components may overfit the data and result in poor interpretations. Many statistical methods have been proposed in the past
few decades by using the information criteria. For example, see Leroux (1992), Roeder and Wasserman (1997), Hennig(2004),
De la Cruz-Mes$\acute{i}$a et al. (2008) and many others. However, these methods are all based on the complete model search
algorithm, which result in heavy computation burden. To improve the computational efficiency, data-driven procedures are much
more preferred. Recently, Chen and Khalili (2008) used the SCAD penalty (Fan and Li, 2001) to penalize the difference of
location parameters for mixtures of univariate location distributions; Kom$\acute{a}$rek and Lesaffre (2008) suggested to
penalize  the reparameterized mixture weights in the generalized mixed model with Gaussian mixtures; Heinzl and Tutz (2014)
constructed a group fused lasso penalty in linear-mixed models; Huang et. al. (2016) proposed a penalized likelihood method
in finite Gaussian mixture models. Most of them are developed for independent data or based on the full likelihood. However,
the full likelihood is often difficult to specify in formulating a mixture model for longitudinal data, particularly for
correlated discrete data.

Instead of specifying the form of distribution of the observations, a quasi-likelihood method (Wedderburn, 1974) gives consistent estimates of parameters in mixture regression models that only needs the relation between the mean and variance of each observation. Inspired by its nice property, in this paper, we propose a new penalized method based on quasi-likelihood for mixture regression models to deal with the above mentioned problems
simultaneously. This would be the first attempt to handle both balanced and unbalanced longitudinal data that only requires
the first two moment conditions of the model distribution. By penalizing the logarithm of mixture proportions, our approach
can simultaneously select the number of mixing components and estimate the mixture proportions and unknown parameters in the
semiparametric mixture regression model. The number of components can be consistently selected. And given the number of
components, the estimators of mixture proportions and regression parameters can be root-$n$ consistent and asymptotically
normal. By taking account of the within-component dispersion, we further develop a modified EM algorithm to improve the
classification accuracy. Simulation results and the application to the motivating PBC data demonstrate the feasibility and
effectiveness of the proposed method.

The rest of the paper is organized as follows. In Section~\ref{sec2}, we introduce a new penalized method for learning
semiparametric mixture regression models with longitudinal data. Section~\ref{sec3} presents the corresponding theoretical
properties and Section~\ref{sec4} provides a modified EM algorithm for implementation. In Section~\ref{sec5}, we assess the
finite sample performance of the proposed method via simulation studies. We apply the proposed method to the PBC data in
Section~\ref{sec6}, and conclude the paper with Section~\ref{sec7}. All technical proofs are provided in Appendix.

\section{Learning semiparametric mixture of regressions}\label{sec2}
\subsection{Model specification}

In a longitudinal study, suppose $Y_{ij}$ is the response variable measured at the $j$th time point for the $i$th subject,
and $X_{ij}$ is the corresponding $p\times 1$ vector of covariates, $i=1,\ldots,n$, $j=1,\ldots,m_i$. Let
$Y_i=(Y_{i1},\ldots,Y_{im_i})^T$ and $X_i=(X_{i1},\ldots,X_{im_i})^T$. In general, the observations for different subjects
are independent, but they may be correlated within the same subject. We assume that the observations of each subject belong
to one of $K$ classes (components) and $u_i \in \{1,\ldots,K\}$ is the corresponding latent class variable. Assume that $u_i$ has a discrete distribution ${\rm{P}}(u_i=k)=\pi_k$, where $\pi_k, k=1,\ldots,K$, are the
positive mixture proportions satisfying $\sum^K_{k=1}\pi_k=1$. Given $u_i=k$ and $X_{ij}$, suppose the conditional mean of
$Y_{ij}$ is
\begin{equation}\label{mean}
\mu_{ijk} \equiv \textup{E}(Y_{ij} \mid X_{ij}, u_i=k) = g(X^T_{ij}\beta_k),
\end{equation}
where $g$ is a known link function, and $\beta_k$ is a $p-$dimensional unknown parameter vector. The corresponding
conditional variance of $Y_{ij}$ is given by
\begin{equation}\label{variance}
\sigma^2_{ijk} \equiv \textup{var}(Y_{ij} \mid X_{ij}, u_i=k) = \phi_k V(\mu_{ijk}),
\end{equation}
where $V$ is a known positive function and $\phi_k$ is a unknown dispersion parameter. In other words, conditioning on $X_{ij}$, the response variable $Y_{ij}$ follows a mixture distribution
\begin{eqnarray*}
Y_{ij}\mid X_{ij} \sim \sum^K_{k=1}\pi_kf_k(Y_{ij}\mid X^T_{ij}\beta_k,\phi_k),
\end{eqnarray*}
where $f_k(Y_{ij}\mid X^T_{ij}\beta_k,\phi_k)$'s are the components' distributions. To avoid identifiability issues, we assume
that $K$ is the smallest integer such that $\pi_k>0$ for $k=1, \cdots, K$, and $(\beta_a, \phi_a)\neq (\beta_b, \phi_b)$
for $1\leq a < b\leq K$. Denote $\theta=(\beta^T_1, \ldots, \beta^T_K, \phi^T,\pi^T)^T$ with $\beta_k=(\beta_{k1},\ldots,\beta_{kp})^T$
and $\pi=(\pi_1,\ldots,\pi_{K-1})^T$, and $\phi=(\phi_1,\ldots,\phi_K)^T$.

Under the working independence correlation, the (log) quasi-likelihood of the $K$-component marginal mixture regression model is
\begin{equation}\label{quasiY}
Q(\theta) = \sum^n_{i=1}\log\left[\sum^K_{k=1} \pi_k \exp\left\{\sum^{m_i}_{j=1}q(g(X^T_{ij}\beta_k);Y_{ij})\right\}\right],
\end{equation}
where function $q(\mu;y)$ (McCullagh and Nelder, 1989) satisfies $\frac{\partial q(\mu;y)}{\partial \mu} = \frac{y-\mu}{V(\mu)}$.
It is known that, for a generalized linear model with independent data, the quasi-likelihood estimator of the regression coefficient has the same asymptotic properties as the maximum likelihood estimator. While for longitudinal data, it is equivalent to the GEE estimator (Liang and Zeger, 1986), which is consistent even when the working correlation structure is misspecified. Therefore, estimation consistency is expected to hold for the $K$-component marginal mixture regression model (\ref{mean})-(\ref{variance}), and this will be validated in Section~\ref{sec3}.

\subsection{Penalized quasi-likelihood method}
For a fixed number of $K$ components, we can maximize the quasi-likelihood function (\ref{quasiY}) by an expectation-maximization (EM) algorithm, which in the E-step computes the posterior probability of the class memberships and in the M-step estimates the mixture proportions and unknown parameters. However, in practice, the number of components is usually unknown and needs to be inferred from the data itself.

For the proposed marginal mixture regression model, the selection of the number of mixing components can be viewed as a model selection problem. Various conventional methods have been proposed based on the likelihood function and some information theoretic criteria. In particular, the Bayesian information criterion (BIC; Schwarz, 1978) is recommended as a useful tool for selecting the number of components (Dasgupta and Raftery, 1998; Fraley and Rafetery, 2002). Therefore, a natural idea is to propose a BIC-type criterion for selecting the number of mixing components, where the likelihood function is replaced by the quasi-likelihood function (\ref{quasiY}). But our simulation experience shows that it couldn't perform as well as the traditional BIC, since (\ref{quasiY}) is no longer a joint density with integral equals to one.

To avoid calculating the normalizing constant, the penalization technique is preferred. By (\ref{quasiY}), intuitively, the $k$th component would be eliminated if $\pi_k=0$. But in implementation of (\ref{quasiY}), the quasi-likelihood function for the complete data $(u_{ik},Y_i, X_i)$ involves $\log{\pi_k}$ rather than $\pi_k$, where $u_{ik}$ denotes the indicator of whether $i$th subject belongs to the $k$th component (see (\ref{complete}) defined in Section~\ref{sec4} for details). Therefore, it is natural to penalize the logarithm of mixture proportions $\log \pi_k$, $k=1,\ldots,K$. Moreover, note that the gradient of $\log \pi_k$ increases very fast when $\pi_k$ is close to zero, and it would dominate the gradient of nonzero $\pi_l>0$. Consequently, the popular $L_q$ types of penalties may not able to set insignificant $\pi_k$ to zero. In the spirit of penalization in Huang et al. (2016), we propose the following penalized quasi-likelihood function
\begin{equation}\label{penQausi}
Q_{\textup{P}}(\theta) = Q(\theta) - n\lambda\sum^K_{k=1}\{\log(\epsilon+\pi_k) - \log(\epsilon)\},
\end{equation}
where $\lambda$ is a tuning parameter and $\epsilon$ is a very small positive constant. Note that $\log(\epsilon+\pi_k) - \log(\epsilon)$ is an increasing function of $\pi_k$ and is shrunk to zero as the mixing proportion $\pi_k$ goes to zero. Therefore, the proposed method (\ref{penQausi}) can simultaneously determine the number of mixture components and estimate mixture proportions and unknown parameters.

{\rema The small constant $\epsilon$ is introduced to ensure the continuity of the objective function when some of mixture proportions are shrunk continuously to zero.}

{\rema The penalty $n\lambda \sum^K_{k=1}\{\log(\epsilon+\pi_k) - \log(\epsilon)\}$ in (\ref{penQausi}) would over penalize large $\pi_k$ and result in a biased estimator. A more general but slightly more complicated approach is to use $n\sum^K_{k=1} \{\log(\epsilon+p_{\lambda}(\pi_k)) - \log(\epsilon)\}$, where $p_{\lambda}(\cdot)$ is a penalty function that gives estimators with sparsity, unbiasedness and continuity as discussed in Fan and Li (2001).}

\section{Asymptotic properties}\label{sec3}

In this section, we first study the asymptotic property of the maximum quasi-likelihood estimator $\widehat{\theta}$ of (\ref{quasiY}) given the number of mixing components. And then, we establish the model selection consistency of the proposed method (\ref{penQausi}) for the general semiparametric marginal mixture regression model (\ref{mean})-(\ref{variance}).

For a fixed number of $K$ components, denote the true value of parameter vector by $\theta_0$. The components of $\theta_0$ are denoted with a subscript, such as $\pi_{0k}$. We assume that the number of subjects $n$ increases to infinity, while the number of observations $\{m_i\}$ is a bounded sequence of positive integers. Let
\begin{equation}\label{Psi}
\Psi(\theta; Y_i|X_i)=\sum^K_{k=1} \pi_k \exp\left\{\sum^{m_i}_{j=1} q(g(X^T_{ij}\beta_k);Y_{ij})\right\}
\end{equation}
and $\psi(\theta; Y_i\mid X_i)=\log(\Psi(\theta; Y_i\mid X_i))$. 

We assume the following regularity conditions to derive the asymptotic properties.

\begin{enumerate}
\item[C1] The function $g(\cdot)$ has two bounded and continuous derivatives.
\item[C2] The random variables $X_{ij}$'s are bounded on the compact support $\mathcal {A}$ uniformly. For $\theta \in \Omega$, the density function of $X^T_{ij}\beta_k$ is positive and satisfies Lipschitz condition of order 1 on $\mathcal {U}_k = \{u=X^T_{ij}\beta_k: X_{ij}\in \mathcal {A}, i=1,\ldots,n, j=1,\ldots,m_i\}$, $k=1,\ldots,K$.
\item[C3] $\Omega$ is compact and $\theta_0$ is an interior point in $\Omega$.
\item[C4] For each $\theta \in \Omega$, $\psi(\theta; Y_i\mid X_i)$ admits third order partial derivatives with respect to $\theta$. And there exist functions $M_l(X_i, Y_i)$, $l=0,1,2,3$ such that for $\theta$ in a neighborhood of $\theta_0$, $|\psi(\theta; Y_i\mid X_i) -  \psi(\theta_0; Y_i\mid X_i)| \leq M_0(X_i, Y_i)$, $|\partial \psi(\theta; Y_i\mid X_i) /  \partial \theta_j | \leq M_1(X_i, Y_i)$, $| \partial^2 \psi(\theta; Y_i\mid X_i) / \partial \theta_j\partial \theta_k | \leq M_2(X_i, Y_i)$, and $|\partial^3 \psi(\theta_0; Y_i\mid X_i) / \partial \theta_j\partial \theta_k\partial \theta_l| \leq M_3(X_i,Y_i)$ with $E\{M_l(X_i,$ $Y_i)\} < \infty$, for all $i=1,\ldots,n$.
\item[C5] $\theta_0$ is the identifiably unique maximizer of $E\{Q(\theta)\}$.
\item[C6] Let $A=\textup{var}\{\partial \psi(\theta_0; Y_i\mid X_i) / \partial \theta\}$. The second derivative matrix $B=\textup{E}\{-\partial^2 \psi(\theta_0; Y_i\mid X_i) / \partial \theta\partial \theta^T\}$ is positive definite.
\end{enumerate}

Conditions C1-C2 are typical assumptions in the estimation literature, which are also found in Xu and Zhu (2012) and Xu et. al. (2016). Conditions C3-C6 are mild conditions in the literature of mixture models, which are used for the proof of weak consistency and asymptotic normality.

{\theo \label{theo1}
Under conditions C1-C6, the maximum quasi-likelihood estimator $\widehat{\theta}$ of (\ref{quasiY}) given the number of components is consistent and has the asymptotic normality
\begin{eqnarray*}
\sqrt{n}(\widehat{\theta} - \theta_0) \stackrel{L}{\longrightarrow} N(0, B^{-1}AB^{-1}).
\end{eqnarray*}}

Next, we study the model selection consistency of the proposed method (\ref{penQausi}) for the marginal mixture regression model (\ref{mean})-(\ref{variance}). We assume that there are $K_0$ mixture components, $K_0 \leq K$ with $\pi_l=0$, for $l=1,\ldots, K-K_0$ and $\pi_l=\pi_{0k}$ for $l=K-K_0+1, \ldots, K$, $k=1,\ldots,K_0$. In the spirit of locally conic parametrization (Dacunha-Castelle and Gassiat, 1997), define $\pi_l=\lambda_l\eta$, $l=1,\ldots,K-K_0$ and $\pi_l=\pi_{0k}+\rho_k\eta$, $l=K-K_0+1, \ldots, K$, $k=1,\ldots,K_0$. Then, the function (\ref{Psi}) can be rewritten as
\begin{eqnarray*}
\Psi(\eta,\gamma;Y_i\mid X_i) = \sum^{K-K_0}_{l=1}\lambda_l\eta f(\beta_l; Y_i\mid X_i)+ \sum^{K_0}_{k=1}(\pi_{0k}+\rho_k\eta) f(\beta_{0k}+\eta\delta_k; Y_i\mid X_i),
\end{eqnarray*}
where $$f(\beta_l; Y_i\mid X_i) = \exp\left\{\sum^{m_i}_{j=1} q(g(X^T_{ij}\beta_l);Y_{ij})\right\}$$ and
$$\gamma = (\lambda_1,\ldots,\lambda_{K-K_0}, \rho_1,\ldots, \rho_{K_0}, \beta^T_1, \ldots,
\beta^T_{K-K_0}, \delta^T_1,\ldots,\delta^T_{K_0})^T$$
with restrictions
$\lambda_l \geq 0, \beta_l \in R^p, l=1,\ldots,K-K_0,$
$\delta_k \in R^p, \rho_k \in R, k=1,\ldots,K_0,$
$\sum^{K-K_0}_{l=1}\lambda_l + \sum^{K_0}_{k=1}\rho_k=0$ and $\sum^{K-K_0}_{l=1}\lambda^2_l + \sum^{K_0}_{k=1}\rho^2_k + \sum^{K_0}_{k=1}\|\delta_k\|^2=1$.
By the permutation, such a parametrization is locally conic and identifiable. And then, the penalized quasi-likelihood function (\ref{penQausi}) can be rewritten as
\begin{equation}\label{conic}
Q_{\textup{P}}(\eta, \gamma) \equiv \sum^n_{i=1}\log\{\Psi(\eta,\gamma;Y_i\mid X_i)\}  - n\lambda\sum^K_{k=1}\{\log(\epsilon+\pi_k) - \log(\epsilon)\}.
\end{equation}

To establish the model selection consistency of the proposed method, we need the following additional conditions:
\begin{enumerate}
\item[C7] There exists a positive constant $\varepsilon$ such that $g(X^T_{ij}\beta_k)$ and $V(g(X^T_{ij}\beta_k))$ are bounded on $\mathcal{B}=\{\beta: \|\beta-\beta_0\|\leq \varepsilon\}$ uniformly in $i=1,\ldots,n$, $j=1,\ldots,m_i$, $k=1,\ldots,K$.
\item[C8] Let $\Sigma_{ik}=\textup{cov}(Y_i\mid X_i, u_i=k)$, and $V_{ik}$ be a $m_i \times m_i$ diagonal matrix with $j$th element $\sigma^2_{ijk}$. Both the eigenvalues of $\Sigma_{ik}$ and $V_{ik}$ are uniformly bounded away from 0 and infinity.
\end{enumerate}

Condition C7 is analogous to conditions (A2) and (A6) in Wang (2011), which is generally satisfied for marginal models. For example, when the marginal model follows a Poisson regression, $V(g(X^T_{ij}\beta_k))=g(X^T_{ij}\beta_k)=\exp(X^T_{ij}\beta_k)$'s
are uniformly bounded around on $\mathcal{B}$. Condition C8 is similar to conditions (C3) and (C4) in Huang et. al. (2007), which ensures the non-singularity of the covariance matrices and the working covariance matrices.

{\theo \label{theo2}
Under conditions C1-C8, if $\lim_{n \rightarrow \infty}\sqrt{n}\lambda = a$ and $\epsilon=o(n^{-1/2}/\log{n})$, where $a$ is a constant, there exists a local maximizer $(\eta, \gamma)$ of (\ref{conic}) such that $\eta=O_p(n^{-1/2})$, and for such local maximizer, we have $\widehat{K}\stackrel{P}{\longrightarrow} K_0$.}

Theorem \ref{theo2} indicates that by choosing an appropriate tuning parameter $\lambda$ and a small constant $\epsilon$, the proposed method (\ref{penQausi}) can select the number of mixing components consistently.

\section{Implementation and tuning parameter selection}\label{sec4}

In this section, we propose an algorithm to implement the proposed method (\ref{penQausi}) and a procedure to select the tuning parameter $\lambda$.

\subsection{Modified EM Algorithm}
Since the membership of each subject is unknown, it is natural to use EM algorithm to implement (\ref{penQausi}). But note that the criterion (\ref{penQausi}) is a function unrelated to different dispersion parameters $\phi_k, k=1,\ldots, K$, therefore, the naive EM algorithm may decrease the classification accuracy for the observations by ignoring the within-component dispersion. Therefore, we here propose a modified EM algorithm in consideration of different component dispersion.

Let $u_{ik}$ denote the indicator of whether the $i$th subject is in the $k$th class. That is, $u_{ik}=1$ if the $i$th subject belongs to the $k$th component, and $u_{ik}=0$ otherwise. If the missing data $\{u_{ik}, i=1,\ldots,n, k=1,\ldots,K\}$ were observed, the penalized quasi-likelihood function for the complete data is given by
\begin{equation}\label{complete}
\sum^n_{i=1}\sum^K_{k=1}u_{ik}\left\{\log{\pi_k} + \sum^{m_i}_{j=1}q(g(X^T_{ij}\beta_k);Y_{ij})\right\} - n\lambda\sum^K_{k=1}\{\log(\epsilon+\pi_k) - \log(\epsilon)\}.
\end{equation}

Denote $\Theta=(\pi^T, \beta^T, \phi^T)^T$ as the vector of all parameters in the $K$-component marginal mixture regression model (\ref{mean})-(\ref{variance}) with $\beta=(\beta^T_1,\ldots,\beta^T_K)^T$. In the E-step, given the current estimate $\Theta^{(t)} = (\pi^{(t)T}, \beta^{(t)T},\phi^{(t)T})^T$, we impute values for the unobserved $u_{ik}$ by
\begin{eqnarray*}
\widehat{u}^{(t+1)}_{ik} = \frac{\pi^{(t)}_k \exp\left\{\sum^{m_i}_{j=1} \tilde{q}(g(X^T_{ij}\beta^{(t)}_k), \phi^{(t)}_k;Y_{ij})\right\}}{\sum^K_{l=1}\pi^{(t)}_l \exp\left\{\sum^{m_l}_{j=1} \tilde{q}(g(X^T_{ij}\beta^{(t)}_l),\phi^{(t)}_l;Y_{ij})\right\}},
\end{eqnarray*}
where $\tilde{q}(\mu,\phi;y) = \int^{\mu}_y \frac{y-t}{\phi V(t)}dt.$
Plugging them into (\ref{complete}), we obtain the function
\begin{equation}\label{expected} \sum^n_{i=1}\sum^K_{k=1}\widehat{u}^{(t+1)}_{ik}\left\{\log{\pi_k} + \sum^{m_i}_{j=1}q(g(X^T_{ij}\beta_k);Y_{ij})\right\}- n\lambda \sum^K_{k=1}\{\log(\epsilon+\pi_k) - \log(\epsilon)\}.
\end{equation}
In the M-step, the goal is to update $\pi^{(t)}$ and $\beta^{(t)}$ by maximizing (\ref{expected}) with the constraint $\sum^K_{k=1}\pi_k=1$ and update $\phi^{(t)}$ by the residual moment method. Specifically, to update $\pi^{(t)}$, we solve the following equations
\begin{eqnarray*}
\frac{\partial}{\partial \pi}\left\{\sum^n_{i=1}\sum^K_{k=1}\widehat{u}^{(t+1)}_{ik} \log{\pi_k} - n\lambda \sum^K_{k=1}\{\log(\epsilon+\pi_k) - \log(\epsilon)\} - \xi(\sum^K_{k=1}\pi_k-1)\right\}=0,
\end{eqnarray*}
where $\xi$ is the Lagrange multiplier. Then, when $\epsilon$ is very close to zero, it gives
\begin{equation}\label{piupdate}
\pi^{(t+1)}_k = \max\left\{0, \frac{1}{1-\lambda K}\left[\frac{1}{n}\sum^n_{i=1}\widehat{u}^{(t+1)}_{ik}-\lambda\right]\right\}, k=1,\ldots, K.
\end{equation}
$\beta^{(t)}_k$ can be updated by solving the following equations
\begin{eqnarray*}
\sum^n_{i=1}\sum^{m_i}_{j=1} \widehat{u}^{(t+1)}_{ik} g'(X^T_{ij}\beta_k)X_{ij} \frac{Y_{ij}-g(X^T_{ij}\beta_k)}{V(g(X^T_{ij}\beta_k))}=0,
\end{eqnarray*}
where $g'(\cdot)$ is the first derivative of $g$, $k=1,\ldots, K$. And using the residual moment method, we update $\phi^{(t)}$ as follows
\begin{eqnarray*}
\phi^{(t+1)}_k = \sum^n_{i=1} \frac{\widehat{u}^{(t+1)}_{ik}}{\sum^n_{i'=1}m_{i'}  \widehat{u}^{(t+1)}_{i'k}} \sum^{m_i}_{j=1}\frac{\{Y_{ij}-g(X^T_{ij}\beta^{(t)}_k)\}^2}{V(g(X^T_{ij}\beta^{(t)}_k))}, k=1,\ldots,K.
\end{eqnarray*}

{\rema In the initial step, we pre-specify a large number of components, and once a mixing proportion is shrunk to zero by (\ref{piupdate}), the corresponding parameters in this component are set to zero and fewer components are kept for the remaining EM iterations. Here we use the same notation $K$ for the whole process.  In practice, during the iterations, $K$ becomes smaller and smaller until the algorithm converges.}

{\rema Although in theory we require $\epsilon=o(n^{-1/2}/\log{n})$, we can update $\pi$ using (\ref{piupdate}) without choosing $\epsilon$ in practice.}

\subsection{Turning Parameter Selection and Classification Rule}

In terms of selecting the tuning parameter $\lambda$, we follow the suggestion in Wang, Li, and Tsai (2007) and use a BIC-type criterion:
\begin{equation}\label{BIC}
\textup{BIC}(\lambda) = -2\sum^n_{i=1} \log\left[\sum^{\widehat{K}}_{k=1} \widehat{\pi}_k \exp\left\{\sum^{m_i}_{j=1} \tilde{q}(g(X^T_{ij}\widehat{\beta}_k), \widehat{\phi}_k;Y_{ij})\right\}
\right] + \widehat{K}(p+2)\log{n},
\end{equation}
where $\widehat{K}$ and $\widehat{\beta}$ are estimators of $K_0$ and $\beta_0$ by maximizing (\ref{penQausi}) for a given $\lambda$.

Let $\widehat{K}$, $\widehat{\pi}$, $\widehat{\beta}$, and $\widehat{\phi}$ be the final estimators of the number of components, the mixture proportions and unknown parameters, respectively. Then, in the sense of clustering, a subject can be assigned to the class whose empirical posterior is the largest. For example, a subject $(Y^*, X^*)$ with $m$ times observations is assigned to the class
\begin{equation}\label{classification}
k^* = \arg \max_{1\leq k \leq \widehat{K}} \widehat{\pi}_k \exp\left\{\sum^m_{j=1}\tilde{q}(g(X^{*T}_{ij}\widehat{\beta}_k), \widehat{\phi}_k;Y^*_{ij})\right\}.
\end{equation}
Consequently, a nature predictor of $Y^*$ is given by $g(X^T \widehat{\beta}_{k^*})$.

{\rema One may claim that $\widehat{\beta}$ would loss some efficiency if the within-subject correlation is strong. It would be better to incorporate correlation information to gain estimation efficiency. However, a correlation analysis would lead to additional computational cost and increase the chance of the convergence problem for the proposed modified EM algorithm. In practice, we suggest to estimate $\beta$ once again given the component information derived from (\ref{penQausi}). Specifically, we first fit the mixture regression model (\ref{mean})-(\ref{variance}) and cluster samples into $\widehat{K}$ classes by (\ref{classification}); then, in each class, the marginal generalized linear model is estimated by applying GEE with a working correlation structure. It is expected that this two-step technique may improve the estimation efficiency if the correlation of the longitudinal data is strong and the working structure is correctly specified.\label{twostage}}

\section{Simulation studies}\label{sec5}

In this section, we conduct a set of Monte Carlo simulation studies to assess the finite sample performance of the proposed method. The maximum initial number of clusters is set to be ten, the initial value for the modified EM algorithm is estimated by K-means clustering and the tuning parameter $\lambda$ is selected by the proposed BIC criterion (\ref{BIC}). To test the classification accuracy and estimation accuracy, we conduct 1000 replications and compare the method (\ref{penQausi}) with the two-step method mentioned in Remark 5 and the QIFC method proposed by Wang and Qu (2014). QIFC is a supervised classification technique for longitudinal data. To permit comparison, we assume that the true number of components, the true class label and the true within-subject correlation are known for the QIFC method. We denote the proposed method and the two-step method as PQL and PQL2 in the following, respectively.

\noindent{\bf {Example 1.}} Motivated by the real data application, we simulate PBC data from a two-component normal mixture as follows. We set $n=300$, $K=2$, $m_i=6$, and $\pi_1=\pi_2=0.5$. For $k$th component, the mean structure of each response is set as
$$E(Y_{ij}) = \beta_{k1}X_{i1} + \beta_{k2}X_{i2} + \beta_{k3}X_{i3} + \beta_{k4}X_{ij4},$$
and the marginal variance is assumed as $\sigma^2_k$. The true values of the regression parameters $\beta_{kj}$'s and the marginal variances $\sigma^2_k$'s are given in Table~\ref{tab2}. Covariates $X_{i1}$ are generated independently from Bernoulli distribution $B(1, 0.5)$ with 0 for placebo and 1 for D-penicillamine. Covariates $X_{i2}$, representing the age of $i$th patient at entry in years, are generated independently from uniform distribution $U(30, 80)$. Covariates $X_{i3}$ are randomly sampled from Bernoulli distribution $B(1, 0.5)$ with 0 for male and 1 for female. For each subject, $m_i=6$ visit times $Z_{ij}$'s are generated, with the first time being equal to 0 and the remaining five visit times being generated from uniform distributions on intervals (350, 390), (710, 770), (1080, 1160), (1450, 1550), and (1820, 1930) days, respectively. Then, let $X_{ij4}=Z_{ij}/30.5$ be the $j$th visit time of $i$th subject in months. Further, for each subject, we assume the within correlation structure is AR(1) with correlation coefficient $0.6$.

To measure the performance of the proposed tuning parameter selector (\ref{BIC}), we show the histograms of the estimated component numbers and report the percentage of selecting the correct number of components. To check the convergence of the proposed modified EM algorithm, we draw the evolution of the penalized quasi-likelihood (\ref{penQausi}) in one run. With respect to classification, we generate 100 new subjects from each component with the same setting as in each configuration and measure the performance in terms of the misclassification error rate. We summarize the median and the 95$\%$ confidence interval of misclassification error rate from a model with correctly identified $K_0$ for PQL and PQL2 and report these quantities for QIFC as well. To measure the performance of the proposed estimators, the mean values of the estimators, the means of the biases, and the mean squared errors (MSE) for the mixture proportions and regression parameters are reported when the number of components $K_0$ is correctly identified. Correspondingly, the mean values, the means of biases, and the MSE of the estimated QIFC estimators are also summarized as a benchmark for comparison. Note that the label switching might arise in practice. Yao (2015) and Zhu and Fan (2016) proposed many feasible labeling methods and algorithms. In our simulation studies, we solve the label switching by putting an order constraint on components' mean parameters.

\begin{figure}
\centering
\subfigure[] {\includegraphics[width=3in]{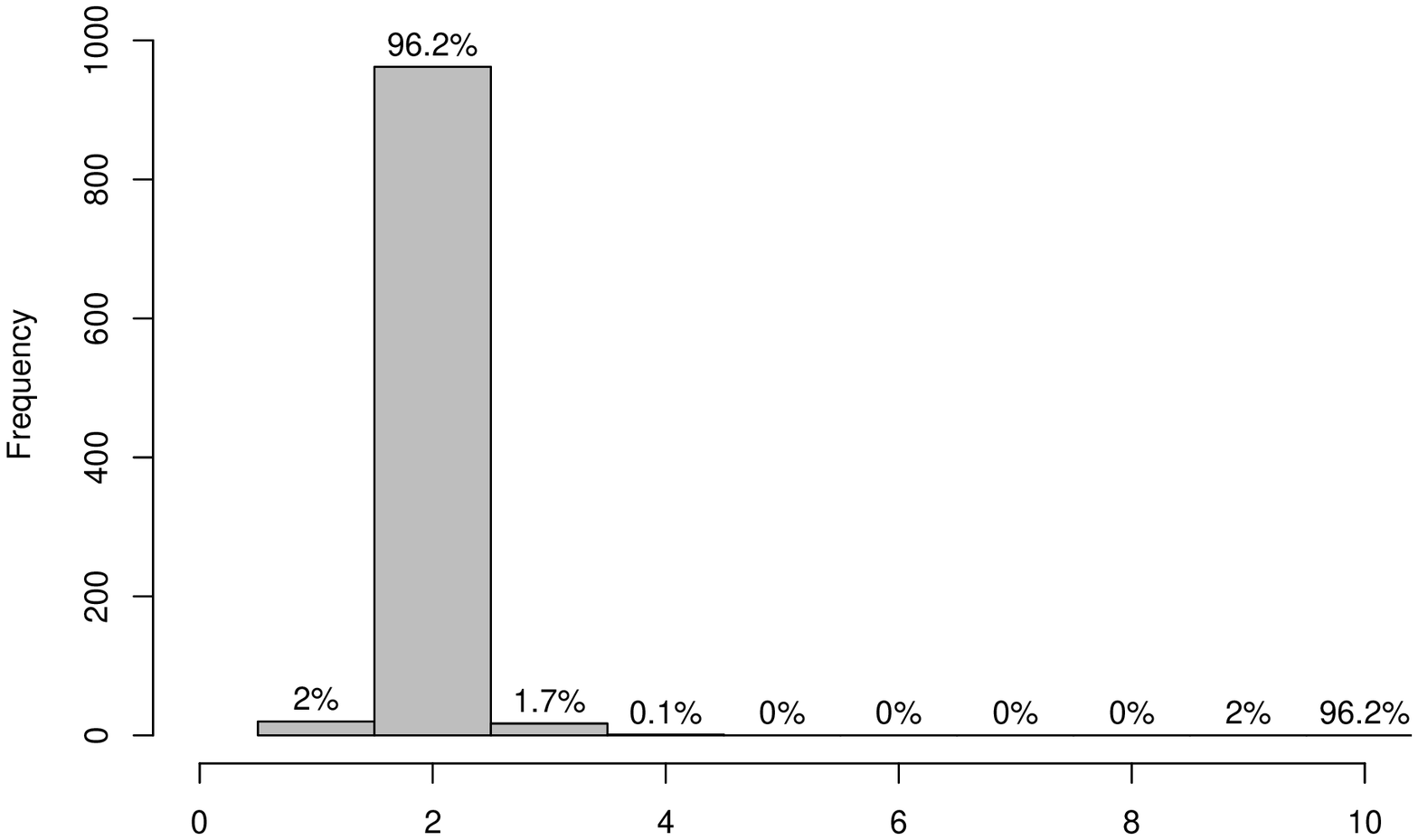}}\quad
\subfigure[]
{\includegraphics[width=2.5in]{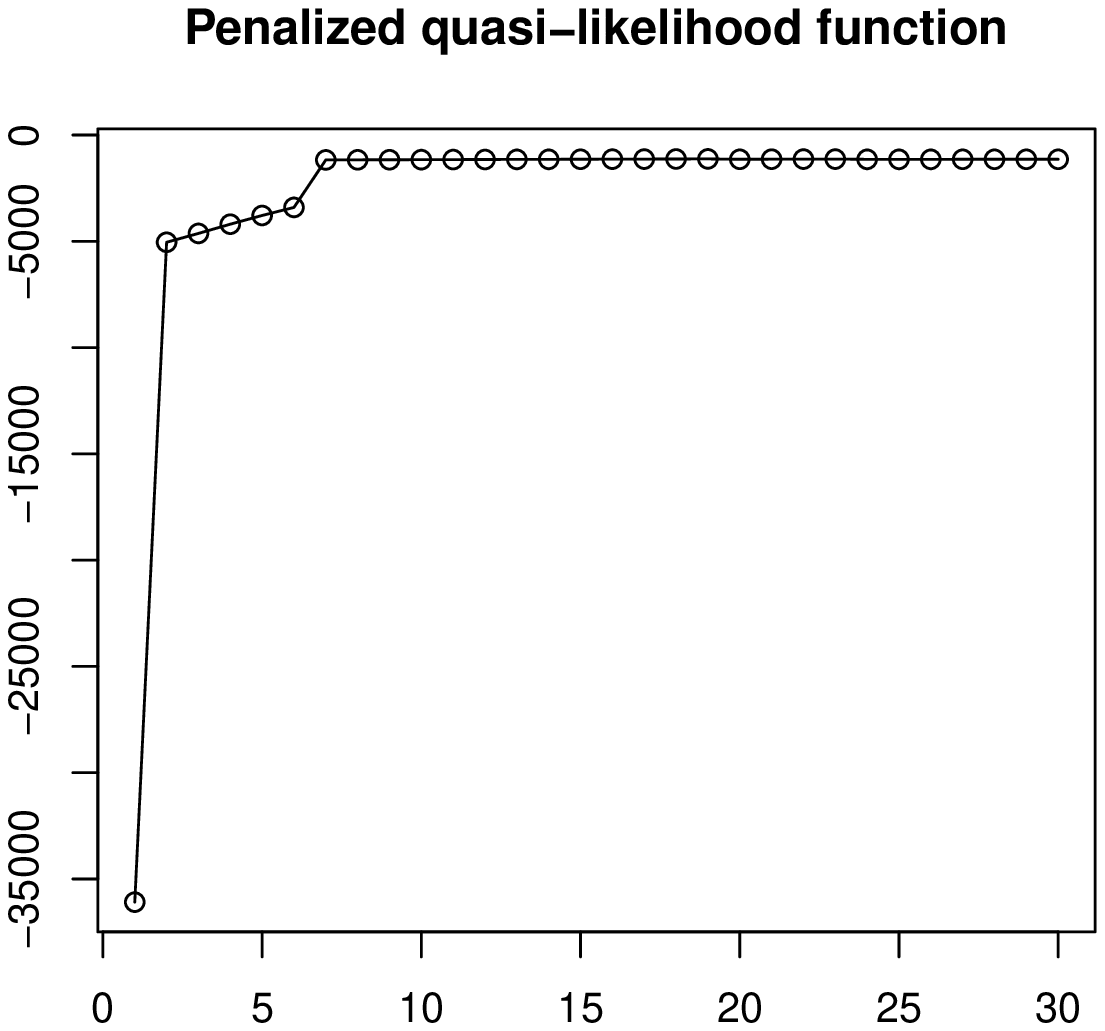}}\quad
\subfigure[]
{\includegraphics[width=3in]{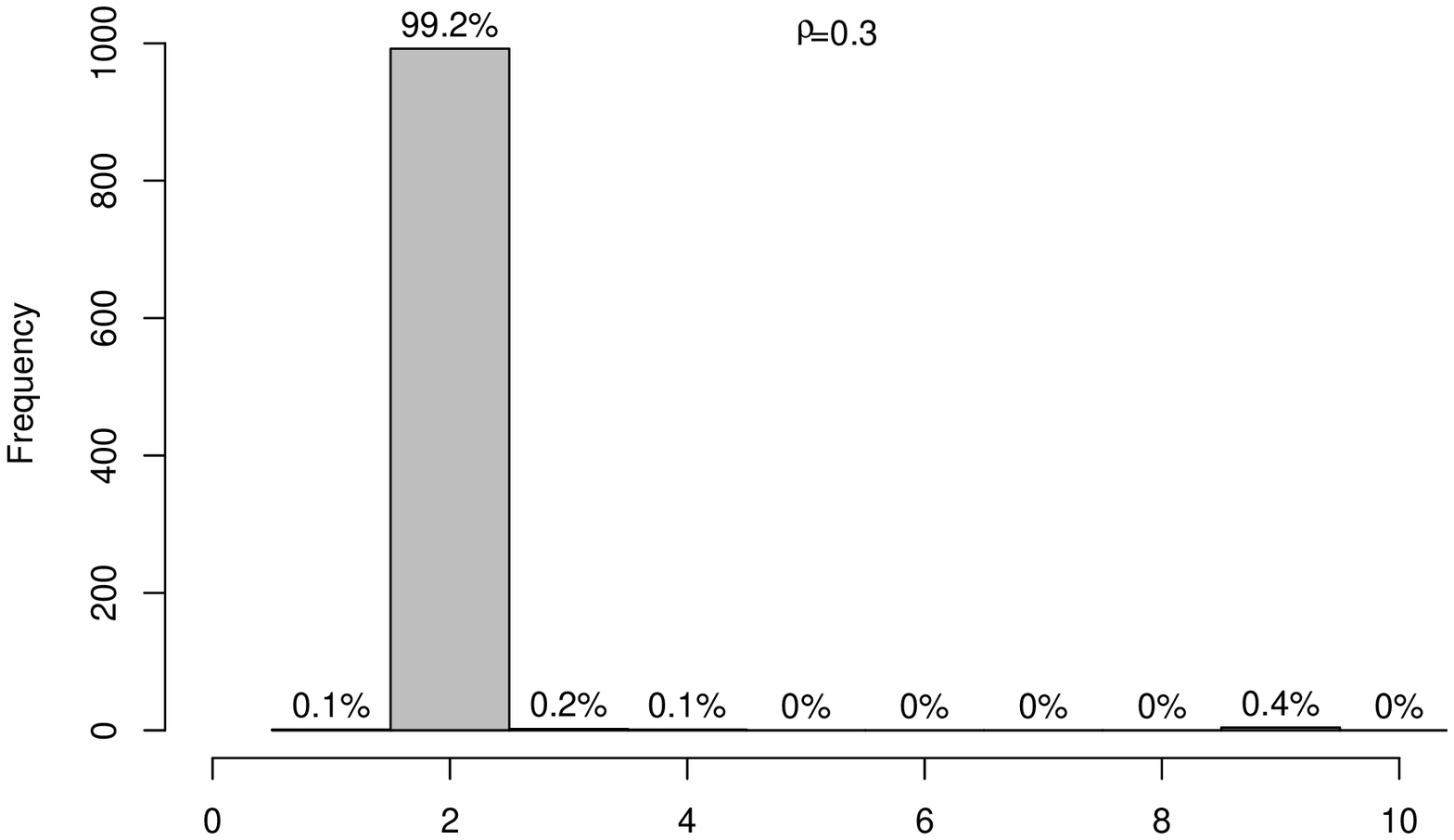}}\quad
\subfigure[] {\includegraphics[width=3in]{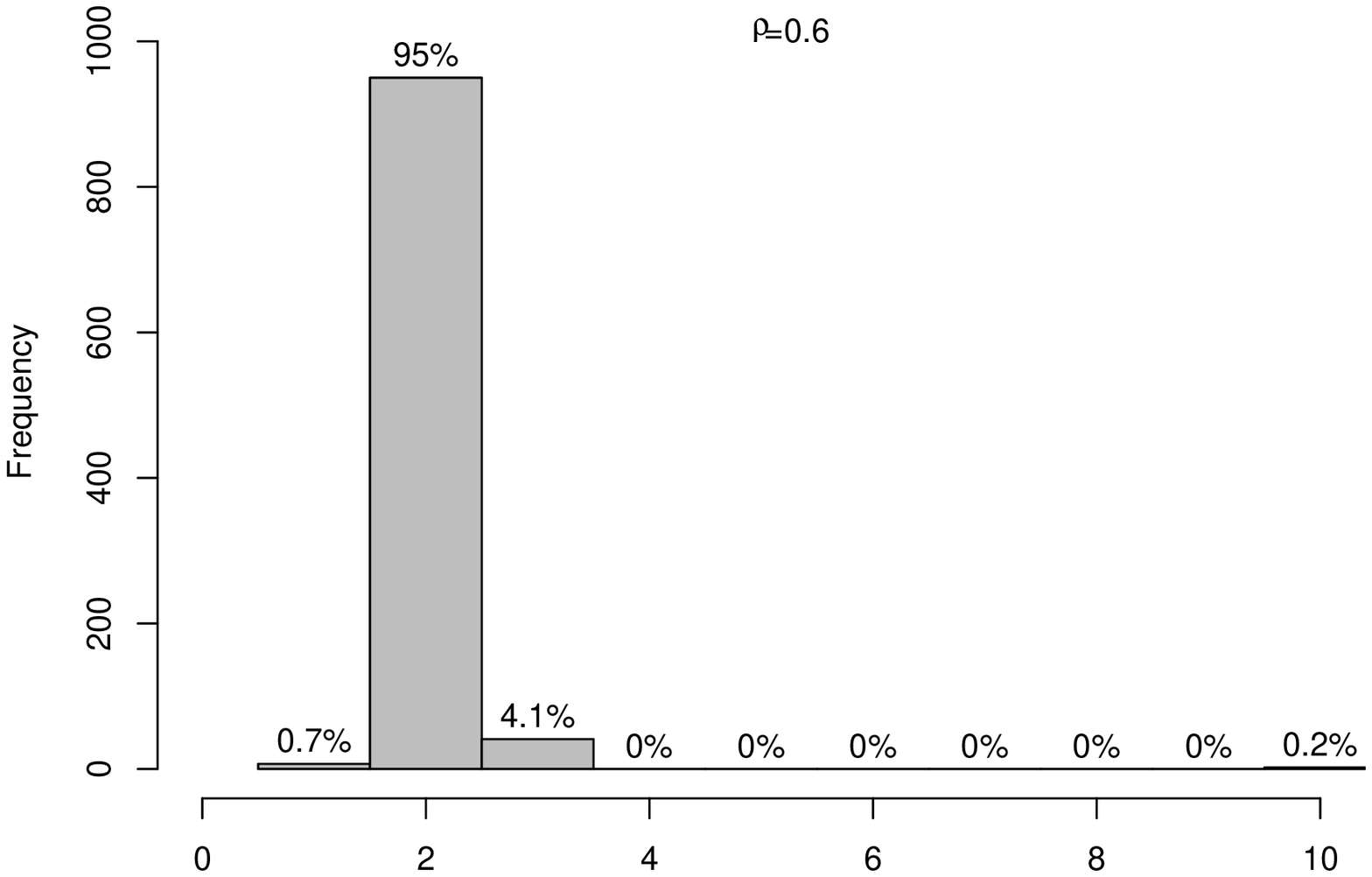}}\quad
\subfigure[] {\includegraphics[width=3in]{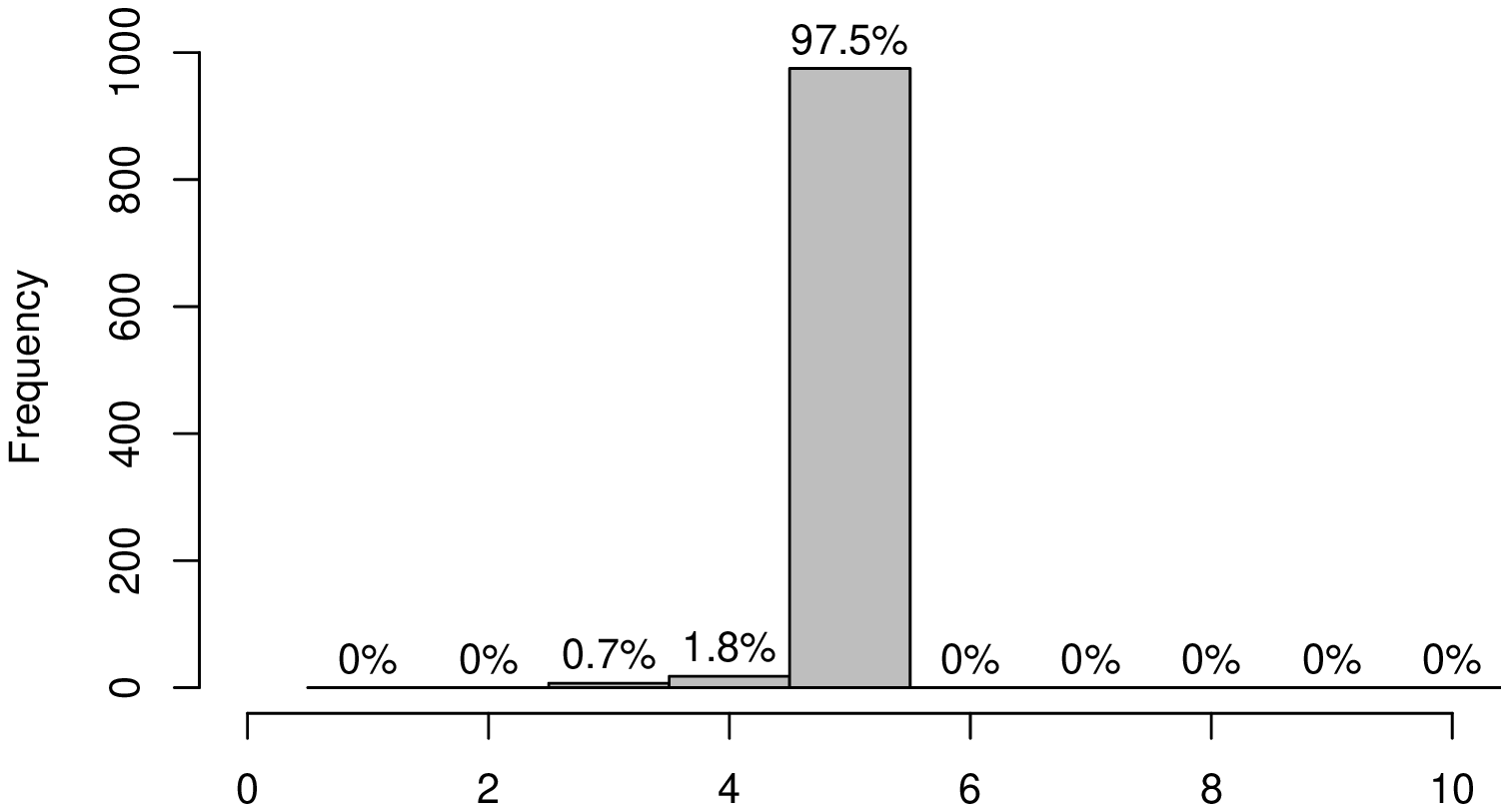}}\quad
\subfigure[]
{\includegraphics[width=3in]{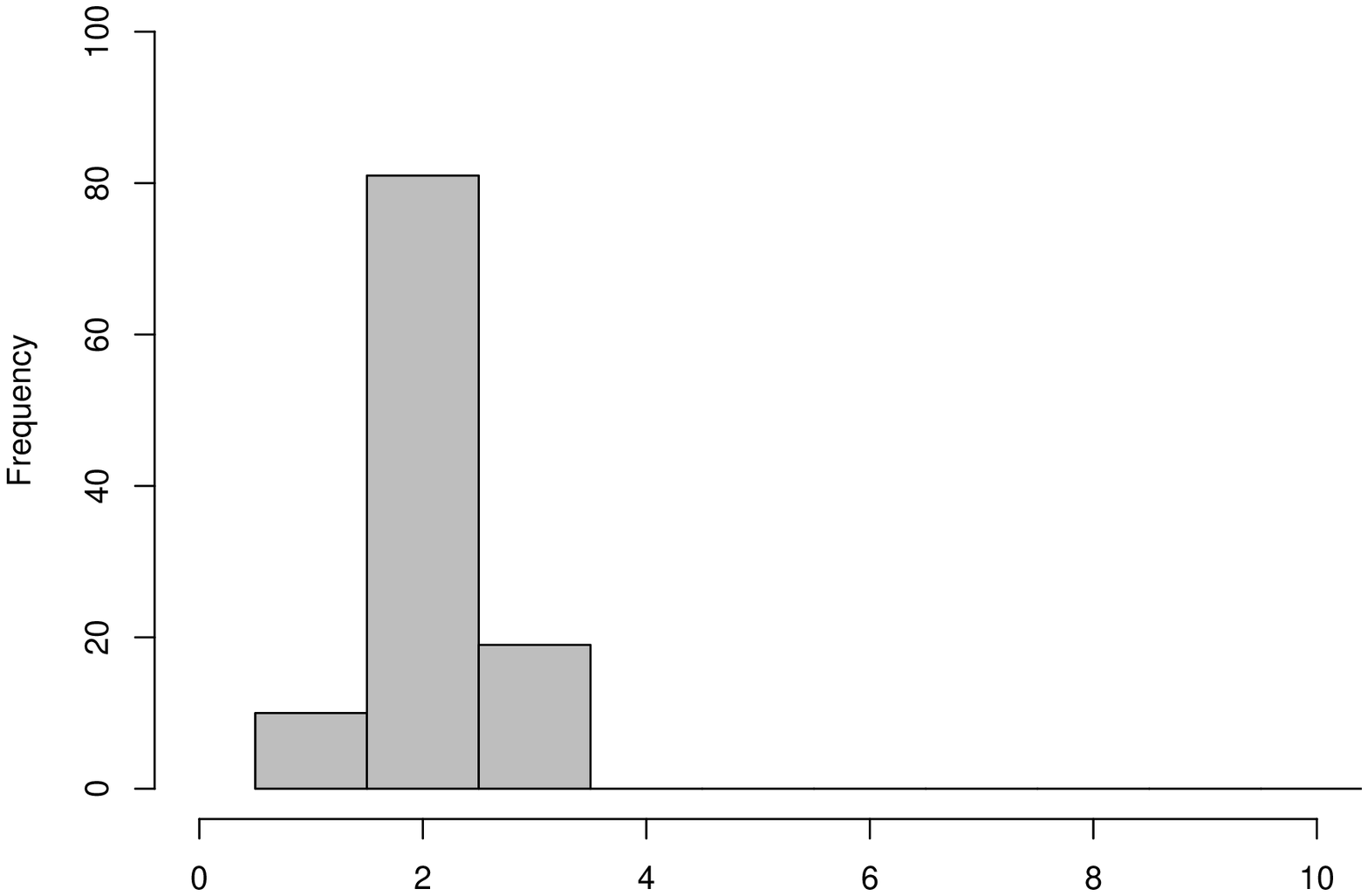}}
\captionsetup{width=1\textwidth}
\caption{Histograms of estimated numbers of components by the proposed PQL method. (a) Example 1, (c) Example 2 with $\rho=0.3$, (d) Example 2 with $\rho=0.6$, (e) Example 3. The value on the top of each bar is the percentage of selecting the corresponding number of components. (b) is the evolution of the penalized quasi-likelihood function for the simulated data set in Example 1 in one typical run. (f) is the histogram of estimated number of components based on 1000 replications in PBC data.}
\label{fig1}
\end{figure}

Figure~\ref{fig1}(a) draws the histogram of the estimated component numbers. It shows that the proposed PQL method with the BIC tuning parameter selector can identify the correct number of components at least with probability 0.962, which is in accordance with the model selection consistency in Theorem~\ref{theo2}. Figure~\ref{fig1}(b) depicts the evolution of the penalized quasi-likelihood function (\ref{penQausi}) for the simulated data set in one run, showing that how our proposed modified EM algorithm converges numerically.

When the number of components is correctly identified, Table~\ref{tab1}(a) reports the median and the 95$\%$ confidence interval of the misclassification error rate from the model-based clustering. We can see that the proposed methods perform better than QIFC with relatively smaller misclassification error rate. Since QIFC is proved asymptotically optimal in terms of misclassification error rate (see Theorem 1 in Wang and Qu, 2014), the observations in Table~\ref{tab1}(a) imply the optimality of the proposed methods in terms of misclassification error rate numerically. Further, in terms of parameter estimation, we summarize the
estimation of mixture proportions and regression parameters in Table~\ref{tab2}. The means of the PQL estimators seem to provide consistent estimates of the regression parameters. It is not surprising that, for regression parameters, the PQL approach performs not as well as the QIFC method with larger bias (in absolute value) and MSE, since QIFC estimators are oracle by assuming the known class memberships and the true within-subject correlation structure. It implies that ignoring the working correlation would affect the efficiency of parameter estimation. However, we can improve the estimating efficiency if the correct correlation information is incorporated. This is reflected in the PQL2 estimators that have much smaller biases (in absolute value) and MSEs compared with the PQL estimators. Indeed, the PQL2 method performs similarly to the QIFC approach.

In addition, combining Table~\ref{tab1}(a) and Table~\ref{tab2}, we can observe that the two-step technique is able to improve the estimation efficiency for the mean regression parameters without reducing the classification accuracy, which validate our guess in Remark 5 numerically. In general, when the within-subject correlation is strong, it is recommended to use PQL2 to provide more predictive power by utilizing the within-subject correlation information.

\begin{center} {
\begin{table}[htb!]
\caption{\label{tab1} {The median and the 95$\%$ confidence interval (CI) of total misclassification error rate in simulation studies. The values of median in Examples 1 and 2 are times 100. For the proposed PQL and PQL2 methods, the results below are summarized based on the models with correctly specified $K_0$ in 1000 replications.}
\hspace{5.5cm}}{{\normalsize \tabcolsep 0.27cm
\renewcommand{\arraystretch}{1.0}
\begin{tabular}{cccccc}
\hline\hline
(a)Example 1 & & Criterion & PQL & PQL2 & QIFC\\
\hline
& & median & 0.000 & 0.000 & 0.058\\
 & & CI & (0.000, 0.000) & (0.000, 0.000) & (0.000, 0.000)\\
\hline\hline
(b)Example 2 & & Criterion & PQL & PQL2 & QIFC\\
\hline
& $\rho=0.3$ & median & 0.234 & 0.232 & 0.235\\
& & CI & (0.000, 0.010) & (0.000, 0.010) & (0.000, 0.010)\\
\hline
& $\rho=0.6$ & median & 0.247 & 0.246 & 0.670\\
& & CI & (0.000, 0.010) & (0.000, 0.010) & (0.000, 0.020)\\
\hline\hline
(c)Example 3 & &¡¡Criterion & PQL & PQL2 & QIFC\\
\hline
& & median & 0.209 & 0.209 & 0.214\\
 & & CI & (0.202, 0.218) & (0.202, 0.218) & (0.204, 0.226)\\
\hline
\end{tabular}
}}
\end{table}
}
\end{center}

\begin{center} {
\begin{table}[htb!]
\caption{\label{tab2} {Estimation results in Example 1: (a) true values of mixture proportions and mixture parameters; (b) means of the parameter estimates; (c) means of the biases for the mixture proportions and mixture parameters; (d) mean squared errors (MSE) for the mixture proportions and mixture parameters. The values of bias and MSE are times 100. For the proposed PQL and PQL2 methods, the results below are summarized based on the models with correctly specified $K_0$ in 1000 replications.}
\hspace{2.5cm}}{{\normalsize \tabcolsep 0.075cm
\renewcommand{\arraystretch}{1.0}
\begin{tabular}{ccccccccccccc}
\hline\hline
Setting &  $\beta_{11}$ & $\beta_{12}$ & $\beta_{13}$ & $\beta_{14}$ & $\beta_{21}$ & $\beta_{22}$ & $\beta_{23}$ & $\beta_{24}$ & $\sigma^2_1$ & $\sigma^2_2$ & $\pi_1$ & $\pi_2$\\
\hline
$K_0=2$ & \multicolumn{12}{c}{True values}\\
 & 0.08 & -0.01 & -0.4 & 0.06 & -0.1 & -0.05 & 3 & 0.3 & 0.5 & 0.8 & 0.5 & 0.5\\
 & \multicolumn{12}{c}{Mean}\\
PQL & 0.079 & -0.010 & -0.402 & 0.060 & -0.099 & -0.050 & 3.005 & 0.300 & 0.511 & 0.789 & 0.500 & 0.500\\
PQL2 & 0.080 & -0.010 & -0.402 & 0.060 & -0.101 & -0.050 & 3.006 & 0.300 & 0.494 & 0.787 & 0.500 & 0.500\\
QIFC & 0.081 & -0.010 & -0.404 & 0.060 & -0.101 & -0.050 & 3.008 & 0.300 & 0.496 & 0.787 & -- & --\\
 & \multicolumn{12}{c}{Bias}\\
PQL & -0.291 & 0.018 & 0.606 & -0.006 & -0.267 & -0.008 & -0.884 & -0.002 & -0.589 & -1.090 & 0.047 & -0.047\\
PQL2 & -0.046 & 0.019 & -0.209 & 0.018 & -0.176 & -0.009 & 0.816 & -0.004 & -0.557 & -0.979 & 0.047 & -0.047\\
QIFC & 0.086 & 0.005 & -0.109 & 0.008 & -0.258 & -0.009 & 0.990 & -0.002 & -0.548 & -0.889 & -- & --\\
 & \multicolumn{12}{c}{MSE}\\
PQL & 0.621 & 0.001 & 4.485 & 0.000 & 1.088 & 0.000 & 3.413 & 0.006 & 2.388 & 0.210 & 0.021 & 0.021\\
PQL2 & 0.608 & 0.002 & 1.374 & 0.002 & 0.921 & 0.001 & 2.095 & 0.006 & 0.887 & 0.274 & 0.021 & 0.021\\
QIFC & 0.558 & 0.001 & 1.281 & 0.000 & 0.949 & 0.001 & 2.131 & 0.002 & 0.098 & 0.253 & -- & --\\
\hline
\end{tabular}
}}
\end{table}
}
\end{center}

\noindent{\bf {Example 2.}} By design, the application of the proposed method is not restricted to continuous responses, and we next evaluate the performance of PQL and PQL2 on count responses. We generate correlated count outcomes from a two-component overdispersed Poisson mixture with mixture proportions $\pi_1=1/3$ and $\pi_2=2/3$. For component 1, the mean function of repeated measurements $Y_{ij}$ is
\begin{eqnarray*}
\log(\mu_{ij1}) = 3X_{ij1}-X_{ij2}+X_{ij3},\quad\quad i=1,\ldots,n_1, j=1,\ldots,m_i,
\end{eqnarray*}
and the marginal variance is $\phi_1\mu_{ij1}= \textup{var}(Y_{ij})= 2\mu_{ij1}$. The correlation structure within a subject is AR(1) with correlation coefficient $\rho$. For component 2, $Y_{i}$ has the same correlation matrix as in component 1, except that
\begin{eqnarray*}
\log(\mu_{ij2}) = 4 - 2X_{ij1}+X_{ij3},\quad\quad i=1,\ldots,n_2, j=1,\ldots,m_i,
\end{eqnarray*}
with dispersion parameter $\phi_2=1$. The number of repeated measurements $m_i$ is randomly drawn from a Poisson distribution with mean 3 and increased by 2, and the sample size is $n=150$. Covariates $X_{ijp}$ are generated independently from uniform distribution $U(0,1)$. Two values of $\rho$ are considered, $\rho=0.3$ and 0.6, to represent different correlation magnitude.

Figure~\ref{fig1}(c) and (d) depict the histograms of the estimated component numbers with different correlation magnitude. It shows that our proposed PQL method can identify the correct model in more than 95$\%$ cases. Even with large within-subject correlation, Figure~\ref{figEx2} (b) in Appendix B shows that the modified EM algorithm converges numerically with the maximum number of components as 10. Once the model is correctly selected, the classification accuracy is quite satisfactory. Table 1(b) implies that PQL and PQL2
provide more predictive power, especially for large within-subject correlation. With respect to bias
and MSE in the estimation of parameters, Table~\ref{tab4} in Appendix B indicates that our modified EM algorithm gives consistent estimates for parameters and mixture proportions by considering the within-class dispersions. Similar to that in Example 1, when the within-subject correlation is large, the PQL2 approach enhances the estimation efficiency by incorporating the correlations within each subject while retaining the class membership prediction accuracy.

\noindent{\bf {Example 3.}} In the third example, we consider a five-component Gaussian mixture of AR(1), exchangeable (CS), and independence (IND). This is a more challenging example with more components but having different correlation structures. Specifically, we generate 500 samples with mixture proportions $\pi_1=\pi_2=0.25$, $\pi_3=\pi_4=0.15$ and $\pi_5=0.2$. Conditional on the class label $u_i$, the response vector $Y_i$ is generated from five multivariate normal distributions:
$$Y_i\mid u_i=k \sim MVN(\beta_{k0}+X_i\beta_k, \sigma^2_kR_{i}^{(k)}), k=1,\cdots,5 $$
where the within-subject correlation structures are set as $R_i^{(1)} =R_{iAR(1)}(0.6)$, $R_i^{(2)}=R_{iAR(1)}(0.6)$,
$R_i^{(3)}=R_{iCS}(0.3)$, $R_i^{(4)}=R_{iCS}(0.3)$, $R_i^{(5)}=R_{iIND}$, and the true values of the regression parameters $(\beta_{k0},\beta_k)$'s and
the variance parameters $\sigma^2_k$'s are given in Table~\ref{tab5}. The number of repeated measurements $m_i$ and the covariates are generated as in Example 2.

Figure~\ref{fig1}(e) draws the histogram of estimated numbers of components and Figure~\ref{figEx3} depicts the evolution of the penalized quasi-likelihood function (\ref{penQausi}) in one run. Though PQL uses a single correlation structure (IND), it is able to identify the correct number of components with high probability, and the corresponding modified EM algorithm converges numerically. Further, the classification results summarized in Table 1(c) shows that PQL gives more accurate prediction of the class's membership compared with QIFC, which is oracle by assuming the known class memberships and the true different within-subject correlation structures. Table~\ref{tab5} in Appendix B indicates, across different finite mixture correlation models, PQL estimators are still consistent. It may loss some efficiency, but can be improved by PQL2.

\section{Application to primary biliary cirrhosis data}\label{sec6}

In this section, we apply the proposed method to study a doubled-blinded randomized trail in primary biliary cirrhosis (PBC) conducted by the Mayo Clinic between 1974 and 1984 (Dickson, Grambsch, Fleming, Fisher, and Langworthy, 1989).

This data set consists of 312 patients who consented to participate in the randomized placebo-controlled trial with D-penicillamine for treating primary biliary cirrhosis until April 1988. Each patient was supposed to have measurements taken at 6 months, 1 year, and annually thereafter. However, 125 of the original 312 patients had died at updating of follow-up in July 1986. Of the remainder, a sizable portion of patients missed their measurements because of worsening medical condition of some labs, which resulted in an unbalanced data structure. A number of variables were recorded for each patient including ID number, time variables such as age and number of months between enrollment and this visit date, categorical variables such as drug, gender and status, continuous measurement variables such as the serum bilirubin level. PBC is a rare but fatal chronic cholestatic liver disease, with a prevalence of about 50-cases-per-millon-population. Affected patients are typically middle-aged women. As in this data set, the sex ratio is 7.2 : 1 (women to men), where the median age of women patients is 49 years old. Identification of PBC is crucial to balancing the need for medical treatment to halt disease progression and extend survival without need for liver transplantation, while minimizing drug-induced toxicities. Biomedical research indicates that serum bilirubin concentration is a primary indicator to help evaluate and track the absence of liver diseases. It is generally normal at diagnosis (0.1$\sim$1 mg/dl) but rise with histological disease progression (Talwalkar and Lindor, 2003). Therefore, we concentrate on modeling the relationship between marker serum bilirubin and other covariates of interest.

We set the log-transformed serum bilirubin level (lbili) as the response variable, since the original level has positive observed values (Murtaugh, Dickson, van Dam, Malinchoc, Grambsch, Langworthy, and Gips, 1994). Figure 2(a) depicts the plot of a set of observed transformed longitudinal profiles of serum bilirubin marker. It shows that the trend of profiles vary over time and the variability may be
large for different patients. The median age of 312
patients is 50 years, but varies between 26 and 79 years. The two sample t-test indicates that there exists significant difference in means of age between male and female groups (p-value = 0.001). Therefore, we consider the marginal semiparametric mixture regression model (\ref{mean})-(\ref{variance}) with the identity link. The mean structure in the $k$th component takes the form
\begin{equation}\label{PBCmean}
\textup{E}(Y_{ij}) = \beta_{k1}\textup{Trt}_{ij} + \beta_{k2} \textup{Age}_{ij} + \beta_{k3} \textup{Sex}_{ij} + \beta_{k4} \textup{Time}_{ij},
\end{equation}
and the marginal variance is assumed as $\textup{var}(Y_{ij}) = \sigma^2_{k}$, $i=1,\ldots, 312$, $j=1,\ldots, m_i$, $k=1,\ldots, K$, where variable $\textup{Trt}$ is a binary variable with 0 for placebo and 1 for D-penicillamine, variable $\textup{Sex}$ is binary with 0 for male and 1 for female, and variable $\textup{Time}$ is the number of months between enrollment and this visit date.

\begin{figure}
\centering
\subfigure[] {\includegraphics[width=4in]{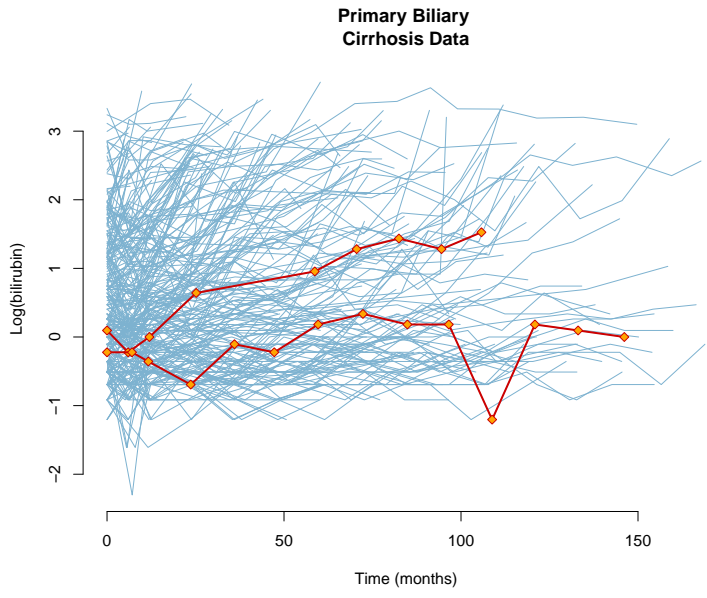}}\quad
\subfigure[] {\includegraphics[width=4in]{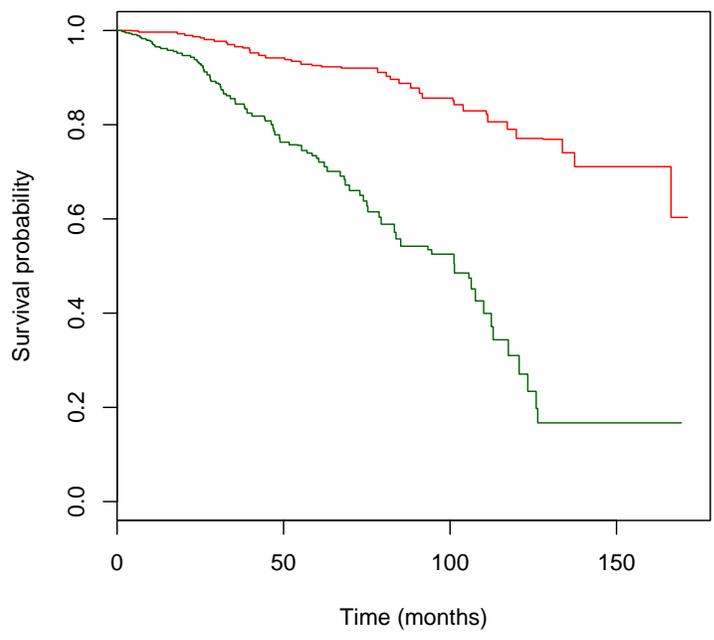}}
\captionsetup{width=1\textwidth}
\caption{(a): Observed transformed longitudinal profiles of serum bilirubin marker. The red lines are profiles of two selected patients (id 2 and 34). (b): The Kaplan-Meier estimate of survival curves for two classes (class 1: red, class 2: green).}
\label{fig2}
\end{figure}

We first standardize data so that there is no intercept term in model (\ref{PBCmean}). Then, we apply the proposed method to simultaneously select the number of components and to estimate the mixture proportions and unknown parameters. As in the simulation studies, the maximum initial number of clusters is set to be ten and the initial value for the modified EM algorithm is estimated by K-means clustering. For comparison purposes, the standard linear mixed-effects model (LMM) with heterogeneity (Verbeke and Lesaffre, 1996; De la Cruz-Mes$\acute{i}$a, Quintana, and Marshall, 2008) is also considered for continuous response variable lbili. The R package mixAK (Kom\'{a}rek and Kom\'{a}rkov\'{a}, 2013) is used to estimate the model and select the number of groups.
The proposed method detects 2 groups, which is same as the clinical classification, while LMM favors 3 groups. Figure~\ref{fig-LMM} in Appendix B depicts the boxplots of residuals in these three groups. The boxplots exhibit the heavy-tailed phenomenon for residuals, especially for those patients in Group 1. It implies that the normality assumptions for the random effects and errors appear inappropriate for modeling this data set. A misspecified distribution of random quantities in the model can seriously influence parameter estimates as well as their standard errors, subsequently leading to invalid statistical inferences. Therefore, it is better to use the proposed semiparametric mixture regression model that only requiring the first two moment conditions of the model distribution. To check the stability of the proposed method, we run our method 100 replications. To be specific, the variable ``status" is a triple variable with 0 for censored, 1 for liver transplanted and 2 for dead. It describes the status of a patient at the endpoint of the cohort study. For each run, we randomly draw 80$\%$ of patients for each of these three status without replacement. Figure~\ref{fig1}(f) shows that our proposed method selects two groups with high probability.

The resulting estimators of parameters and mixture proportions along with the standard deviations are shown in Table~\ref{tab7}. One scientific question of this cohort study is that whether the drug D-pencillamine has effective impact on slowing the rate of increase in serum bilirubin level. According to the estimates and standard deviations with respect to covariate ``Trt" in Table~\ref{tab7}, it implies that there is little benefit of D-pencillamine to lowering the rate of increase in serum bilirubin level even harmful effect, which is in accordance with findings in other literatures (eg Hoofnagle, David, Schafer, Peters, Avigan, Pappas, Hanson, Minuk, Dusheiko, and Campbell, 1986; Pontecorvo, Levinson, and Roth, 1992).

Another goal of this study is to identify groups of patients with similar characteristics by using the values of the marker serum bilirubin and to see how the bilirubin levels evolve over time.
Figure~\ref{fig-PQL} in Appendix B depicts the fitted mean profiles in identified two groups, showing the increasing trend of bilirubin levels in both groups. According to the estimates and standard deviations of parameters in Table~\ref{tab7}, it implies that the covariate ``Time" is significant and bilirubin level increases over time in both treatment and control arms. Moreover, note that $\widehat{\beta}_{14}=0.068 < \widehat{\beta}_{24}=0.313$, which implies that the bilirubin level increases more slowly over time in Group 0. Therefore, from the clinical point of view, Group 0 should correspond to patients with a better prognosis compared to Group 1. To confirm this conclusion, Kaplan-Meier estimates of the survival probabilities are calculated based on data from patients classified in each group. We can see that, from Figure 2(b), the survival prognosis of Group 0 is indeed much better than that of Group 1 with the estimated 5-year survival probability in Group 0 of 0.926 compared to 0.729 in Group 1, and the 10-year survival probabilities 0.771 and 0.310 in Groups 0 and 1, respectively. The p-value of the log rank test is near 0, which implies that the survival distributions corresponding to identified groups are quite different. Further, according to the variable ``status", the group levels for 312 patients are predefined. At the endpoint of the cohort study, 140 of the patients had died, Group 1, while 172 were known to be alive, Group 0. Therefore, it is of interest to compare the classification results using the fitted semiparametric two-component mixture models shown in Table~\ref{tab8}. For comparison purposes, the fitting results and classification results of the QIFC method are presented in Tables~\ref{tab7} and ~\ref{tab8}, respectively. It can be observed that the proposed method provides more accurate classification performance than the QIFC.

\begin{center}{
\begin{table}[htb!]
\caption{\label{tab7} {Parameter estimates for primary biliary cirrhosis data.}
\hspace{5cm}}{{\normalsize \tabcolsep 0.64cm
\renewcommand{\arraystretch}{1.0}
\begin{tabular}{ccccc}
\hline\hline
 & \multicolumn{2}{c}{PQL} & \multicolumn{2}{c}{QIFC}\\
Parameters & Group 0 & Group 1 & Group 0 & Group 1\\
\hline
mixture proportions & 0.512 & 0.487 & -- & --\\
 & (0.129) & (0.129) & -- & --\\
Trt & 0.084 & -0.097 & 0.055 & -0.076\\
 & (0.183) & (0.534) & (0.037) & (-0.113)\\
Age & -0.016 & -0.051 & -0.272 & 0.104\\
 & (0.075) & (0.418) & (-0.261) & (0.262)\\
Sex & -0.366 & 3.220 & -0.125 & -0.204\\
 & (0.097) & (1.219) & (-0.113) & (-0.064)\\
Time & 0.068 & 0.313 & 0.093 & -0.106\\
 & (0.029) & (0.113) & (0.119) & (-0.108)\\
$\sigma^2$ & 0.523 & 0.781 & 0.832 & 2.641\\
 & (0.289) & (0.146) & (0.833) & (2.689)\\
\hline
\end{tabular}
}}
\end{table}
}
\end{center}

\begin{center}{
\begin{table}[htb!]
\caption{\label{tab8} {Agreements and differences between the clinical and model classifications using the PQL and QIFC methods.}
\hspace{5cm}}{{\normalsize \tabcolsep 0.74cm
\renewcommand{\arraystretch}{1.0}
\begin{tabular}{ccccccc}
\hline\hline
 & & \multicolumn{2}{c}{PQL} & \multicolumn{2}{c}{QIFC} & Total\\
\multicolumn{2}{c}{Classify to} & 0 & 1 & 0 & 1 & \\
True & Group 0 & 118 & 54 & 69 & 103 & 172\\
     & Group 1 & 42  & 98 & 21 & 119 & 140\\
     & Total   & 160 & 152 & 90 & 222 & 312\\
\hline
\end{tabular}
}}
\end{table}
}
\end{center}

\section{Conclusion}\label{sec7}

In this paper, we have proposed a penalized method for learning mixture regression models from longitudinal data which is able to select the number of components in an unsupervised way. The proposed method only requires the first two moment conditions of the model distribution, and thus is suitable for both the continuous and discrete responses. It penalizes the logarithm of mixing
proportions, which allows one to simultaneously select the number of components and to estimate the mixture proportions and unknown parameters. Theoretically, we have shown that our proposed approach can select the number of components consistently for general marginal semiparametric mixture regression models. And given the number of components, the estimators of mixture proportions and regression parameters are root-$n$ consistent and asymptotically normal.

To improve the classification accuracy, a modified EM algorithm has been proposed by considering the within-component dispersion. Simulation results and the real data analysis have shown its convergence, but further theoretical investigation is needed. And we have introduced a BIC-type method to select the tuning parameter automatically. Numerical studies show it works well, while the theoretical consistency deserves a further study.

Another issue is the consideration of the within-subject correlation. The proposed penalized approach is introduced under the working independence correlation. Simulation results have implied that it may lose some estimation efficiency, especially when the within-subject correlation is large. Therefore, we suggest a two-step technique to refine the estimates. Simulations show that the efficiency improvement is significant if the correlation information is incorporated and the working structure is correctly specified. It would be worthwhile to systematically study the unsupervised learning of mixtures by incorporating correlations.

Finally, in the presence of missing data at some time points, our implicit assumption is missing completely at random, under which the quasi-likelihood method yield consistent estimates (Liang and Zeger, 1986). Such an assumption is applicable to our motivating example, as patients missed their measurements due to administrative reasons. However, when the missing values are informative, the proposed method has to be modified so as to incorporate missing mechanisms. This is beyond the current scope of the work and would warrant further investigations.

\newpage
\section*{Appendix}

\subsection*{A. Proofs of Theorems}

\noindent {\bf{Proof of Theorem 1}}

Recall $\Psi(\theta; Y_i|X_i)=\sum^K_{k=1} \pi_k \exp\left\{\sum^{m_i}_{j=1} q(g(X^T_{ij}\beta_k);Y_{ij})\right\}$,
and $\psi(\theta; Y_i\mid X_i)=\log(\Psi(\theta; Y_i\mid X_i))$. Under Condition C5, $\theta_0$ is a maximizer of $n^{-1}\sum^n_{i=1}E\{\psi(\theta; Y_i\mid X_i)-\psi(\theta_0; Y_i\mid X_i)\}$. Then, $\theta_0$ is identifiability unique. Therefore, in the spirit of Theorem 17 in Ferguson (1996) and Theorem 2.1 in Bollerslev and Wooldridge (1992), $\widehat{\theta}$ is weak consistent under Conditions C1-C5. Let $\widehat{\theta}^*=\sqrt{n}(\widehat{\theta}-\theta_0)$. Then, $\widehat{\theta}^*$ maximizes
\begin{eqnarray*}
Q_n(\widehat{\theta}^*) = \sum^n_{i=1}\{\psi(n^{-1/2}\widehat{\theta}^*+\theta_0; Y_i\mid X_i) - \psi(\theta_0; Y_i\mid X_i)\}.
\end{eqnarray*}
An application of Taylor expansion yields that
\begin{eqnarray}\label{taylor}
Q_n(\widehat{\theta}^*)&=&\frac{1}{\sqrt{n}}\sum^n_{i=1} \frac{\partial \psi(\theta_0; Y_i\mid X_i)}{\partial \theta}\widehat{\theta}^* + \frac{1}{2}\widehat{\theta}^{*T} \left\{\frac{1}{n}\sum^n_{i=1}\frac{\partial^2 \psi(\theta_0; Y_i\mid X_i)}{\partial \theta \partial \theta^T}\right\} \widehat{\theta}^{*} + o_p(\mathbf{1})\nonumber\\
&\equiv& D_n\widehat{\theta}^{*} + \frac{1}{2}\widehat{\theta}^{*T}B_n\widehat{\theta}^* + o_p(\mathbf{1}),
\end{eqnarray}
where $\mathbf{1}$ is a $pK+(K-1)$ dimensional all-ones vector. It can be shown that $B_n \stackrel{P}{\longrightarrow} -B$. Then, by (\ref{taylor}) and quadratic approximation lemma, we have
$\widehat{\theta}^{*} = B^{-1}D_n + o_p(\mathbf{1})$.
Note that $\textup{var}(D_n) = A$. And under the regularity conditions, we have $D_n \stackrel{L}{\longrightarrow} N(0, A)$. Hence,
\begin{eqnarray*}
\widehat{\theta}^{*} \stackrel{L}{\longrightarrow} N(0, B^{-1}AB^{-1}).
\end{eqnarray*}

In order to establish Theorem 2, we need the following lemma first, which can be derived using similar arguments as the proof of Proposition A.1 of Huang et al. (2016).

For a data pair $(Y,X)$ with $m$ times observations, define $$\Psi_0(Y\mid X) = \sum^{K_0}_{k=1}\pi_{0k} f(\beta_{0k}; Y\mid X).$$
Let $\mathcal{D}$ be the subset of functions of form
\begin{eqnarray*}
d(\gamma;Y\mid X) = \sum^{K_0}_{k=1}\pi_{0k}\sum^{p}_{j=1}\delta_{kj} \frac{D^1_jf(\beta_{0k}; Y\mid X)}{\Psi_0(Y\mid X)} + \sum^{K-K_0}_{l=1} \lambda_l \frac{f(\beta_{l}; Y\mid X)}{\Psi_0(Y\mid X)} + \sum^{K_0}_{k=1}\rho_k \frac{f(\beta_{0k}; Y\mid X)}{\Psi_0(Y\mid X)},
\end{eqnarray*}
where $D^1_jf(\beta_{0k}; Y\mid X)$ is the first derivative of $f(\beta_{0k}; Y\mid X)$ for the $j$th component of $\beta_{0k}$.

{\lemm\label{lemm}
Under conditions C1-C6, $\mathcal{D}$ is a Donsker class.}

{\noindent \it{Proof:}} Under conditions C1-C6, it is straightforward to show that $\mathcal{D}$ satisfies conditions P0 and P1 in Dacunha-Castelle and Gassiat (1999) as in Keribin (2000). Then, there exists a $\Psi_0\nu$-square integrable envelope function $\bar{d}(\cdot)$ such that $|d(\gamma; Y|X)| \leq \bar{d}(Y|X)$. On the other hand, the sequences of coefficients of $d(\gamma; Y|X)$ are bounded under the restrictions imposed on $\gamma$. Hence, similar to the proof of Proposition 3.1 in Dacunha-Castelle and Gassiat (1999), we can show that $\mathcal{D}$ has the Donsker property with the bracketing number $N(\varepsilon)=\varepsilon^{-pK}$.

\noindent{\bf{Proof of Theorem 2}}

In the spirit of proof of Theorem 3.2 in Huang et al. (2016), we divide our proof into two parts. First, we show that there exists a maximizer $(\eta, \gamma)$ such that $\eta=O_p(n^{-1/2})$ when $\lambda=a/\sqrt{n}$. It is sufficient to show that, for a large constant $C$, $Q_{\textup{P}}(\eta, \gamma) < Q_{\textup{P}}(0, \gamma)$ when $\eta=C/\sqrt{n}$. Let $\eta=C/\sqrt{n}$, and note that
\begin{eqnarray*}
& & Q_{\textup{P}}(\eta, \gamma) - Q_{\textup{P}}(0, \gamma) = \sum^n_{i=1}\left\{\log\Psi(\eta,\gamma;Y_i\mid X_i) - \log\Psi_0(Y_i\mid X_i)\right\}\\
& & \qquad - n\lambda\sum^{K}_{l=K-K_0+1}\{\log(\epsilon+\pi_l) - \log(\epsilon+\pi_{0(l-K+K_0)})\}  - n\lambda\sum^{K-K_0}_{k=1}\{\log(\epsilon+\pi_k) - \log(\epsilon)\}.
\end{eqnarray*}
Then,
\vspace{-1.1cm}
\begin{eqnarray*}
& & Q_{\textup{P}}(\eta, \gamma) - Q_{\textup{P}}(0, \gamma) \leq  \sum^n_{i=1}\left\{\log\Psi(\eta,\gamma;Y_i\mid X_i) - \log\Psi_0(Y_i\mid X_i)\right\}\\
& & \qquad \qquad  - n\lambda\sum^{K}_{l=K-K_0+1}\{\log(\epsilon+\pi_l) - \log(\epsilon+\pi_{0(l-K+K_0)})\} := S_1 + S_2.
\end{eqnarray*}
For $S_1$, an application of Taylor expansion yields
\begin{eqnarray*}
S_1 &=& \sum^n_{i=1}\frac{\Psi(\eta,\gamma;Y_i\mid X_i)-\Psi_0(Y_i\mid X_i)}{\Psi_0(Y_i\mid X_i)} - \frac{1}{2}\sum^n_{i=1}\left\{\frac{\Psi(\eta,\gamma;Y_i\mid X_i)-\Psi_0(Y_i\mid X_i)}{\Psi_0(Y_i\mid X_i)}\right\}^2\\
& & + \frac{1}{3}\sum^n_{i=1}t_i\left\{\frac{\Psi(\eta,\gamma;Y_i\mid X_i)-\Psi_0(Y_i\mid X_i)}{\Psi_0(Y_i\mid X_i)}\right\}^3
\end{eqnarray*}
for $\eta=C/\sqrt{n}$, where $|t_i| \leq 1$. By Taylor expansion again for $\Psi(\eta,\gamma;Y\mid X)$ at $\eta=0$, we have
$\Psi(\eta,\gamma;Y\mid X) = \Psi_0(Y\mid X) + \eta\Psi'(0,\gamma;Y\mid X)+\frac{\eta^2}{2}
\Psi''(\tilde{\theta},\gamma;Y\mid X),$
for a $\tilde{\theta} \leq \theta$. Then, by conditions C1-C5, we have
\begin{eqnarray*}
S_1=\left[\sum^n_{i=1}\eta\frac{\Psi'(0,\gamma;Y_i\mid X_i)}{\Psi_0(Y_i\mid X_i)} - \frac{1}{2}\sum^n_{i=1}\eta^2 \left\{\frac{\Psi'(0,\gamma;Y_i\mid X_i)}{\Psi_0(Y_i\mid X_i)}\right\}^2\right](1+o_p(1)).
\end{eqnarray*}
By Lemma~\ref{lemm} for the class $\mathcal{D}$, we know that $\frac{1}{\sqrt{n}}\sum^n_{i=1}\frac{\Psi'(0,\gamma;Y_i\mid X_i)}{\Psi_0(Y_i\mid X_i)}$ converges uniformly in distribution to a Gaussian process and $\sum^n_{i=1}\left\{\frac{\Psi'(0,\gamma;Y_i\mid X_i)}{\Psi_0(Y_i\mid X_i)}\right\}^2=O_p(n)$ by the law of large numbers. Therefore,
\begin{eqnarray*}
S_1=\frac{C}{\sqrt{n}}O_p(\sqrt{n}) - \frac{C^2}{n}O_p(n).
\end{eqnarray*}
For $S_2$, we know that $
|\pi_l-\pi_{0(l-K+K_0)}| = |\eta\rho_{l-K+K_0}| \leq \frac{C}{\sqrt{n}},\quad l=K-K_0+1,\ldots,K,$
by the restriction condition on $\rho_k$, $k=1,\ldots,K_0$. Thus, by Taylor expansion, we have
$$
|S_2| = \left|n\lambda \sum^K_{l=K-K_0+1} \frac{\pi_l-\pi_{0(l-K+K_0)}}{\epsilon+\pi_{0(l-K+K_0)}}\{1+o(1)\}\right|\\
= O(\sqrt{n})\frac{CK_0}{\sqrt{n}}\{1+o(1)\} = O(C),
$$
if $\sqrt{n}\lambda \rightarrow a$. Therefore, when $C$ is large enough, the second term in $S_1$ dominates $S_2$ and other terms in $S_1$. Consequently, we have
$Q_{\textup{P}}(\eta, \gamma) - Q_{\textup{P}}(0, \gamma) < 0$
with probability tending to one. Hence, there exists a maximizer $(\eta,\gamma)$ such that $\eta=O_p(n^{-1/2})$ with probability tending to one.

Then, we show that $\widehat{K}=K_0$ or $\widehat{\pi}_k=0, k=1,\ldots,K-K_0$ when the maximizer $(\eta,\gamma)$ satisfies $\eta=O_p(n^{-1/2})$. We first show that, for any maximizer $Q_{\textup{P}}(\eta^*, \gamma^*)$ with $|\eta^*| \leq Cn^{-1/2}$, if there is a $k\leq K - K_0$ such that $Cn^{-1/2} \geq \pi^*_k > n^{-1/2}/\log n$, there exists another maximizer of $Q_{\textup{P}}(\eta, \gamma)$ in the area of $|\eta| \leq Cn^{-1/2}$. It is equivalent to show that $Q_{\textup{P}}(\eta^*, \gamma^*) < Q_{\textup{P}}(0, \gamma^*)$ holds with probability tending to one for any such kind maximizer $Q_{\textup{P}}(\eta^*, \gamma^*)$ with $|\eta^*| \leq Cn^{-1/2}$. For any $k < K-K_0+1$, we have
\begin{eqnarray*}
& & Q_{\textup{P}}(\eta^*, \gamma^*) - Q_{\textup{P}}(0, \gamma^*) \leq \sum^n_{i=1}\left\{\log\Psi(\eta^*,\gamma^*;Y_i\mid X_i) - \log\Psi_0(Y_i\mid X_i)\right\}\\
& & - n\lambda\sum^{K}_{l=K-K_0+1}\{\log(\epsilon+\pi^*_l) - \log(\epsilon+\pi_{0(l-K+K_0)})\} - n\lambda \{\log(\epsilon+\pi^*_k) - \log\epsilon\}\\
&:=& S_1 + S_2 + S_3.
\end{eqnarray*}
As shown before, we have $S_1+S_2=O_p(C^2)$. For $S_3$, because $\epsilon=o(n^{-1/2}/\log n)$ and $\pi_k < n^{-1/2}/\log n$, we have
\begin{eqnarray*}
|S_3| = O(n\cdot \frac{C}{\sqrt{n}})\log \frac{\pi^*_k}{\epsilon} = O(n^{1/2}),
\end{eqnarray*}
which implies that $S_3$ dominates $S_1$ and $S_2$, and hence $Q_{\textup{P}}(\eta^*, \gamma^*) < Q_{\textup{P}}(0, \gamma^*)$. So, in the following step, we only need to consider the maximizer $Q_{\textup{P}}(\widehat{\eta}, \widehat{\gamma})$ with $|\widehat{\eta}| \leq Cn^{-1/2}$ and $\widehat{\pi}_k < n^{-1/2}/\log n$ for $k \leq K-K_0$.

Let $Q^*(\theta)=Q_{\textup{P}}(\theta) - \xi (\sum^K_{k=1}\pi_k - 1)$, where $\xi$ is a Lagrange multiplier. Then, it is sufficient to show that,  for the maximizer $(\eta, \gamma)$,
\begin{equation}\label{eq1}
\frac{\partial Q^*(\theta)}{\partial \widehat{\pi}_k} < 0 \quad \mbox{for}\quad \widehat{\pi}_k < \frac{1}{\sqrt{n}\log n}, k \leq K-K_0
\end{equation}
with probability tending to one. For $k=1,\ldots,K$, note that $\widehat{\pi}_k$ satisfies
\begin{equation}\label{eq2}
\frac{\partial Q^*(\theta)}{\partial \widehat{\pi}_k}=\sum^n_{i=1}\frac{f_k(\beta_k;Y_i\mid X_i)}{\sum^K_{l=1} \widehat{\pi}_l f_l(\beta_l;Y_i\mid X_i)} - n\lambda \frac{1}{\epsilon+\widehat{\pi}_k} - \xi = 0,
\end{equation}
where $f_l(\beta_l;Y_i\mid X_i) = \exp\left\{\sum^{m_i}_{j=1}q(g(X^T_{ij}\beta_l);Y_{ij})\right\}$. By the law of large numbers, the first term of (\ref{eq2}) is of order $O_p(n)$. If $k > K-K_0$ and $\eta=O_p(n^{-1/2})$, we have that
\begin{eqnarray*}
\widehat{\pi}_k = \pi_{0(k-K+K_0)}+O_p(n^{-1/2}) > \frac{1}{2}\min\{\pi_{01},\ldots,\pi_{0K_0}\}.
\end{eqnarray*}
Hence, the second term of (\ref{eq2}) is of order $O_p(n\lambda) = o_p(n)$. Thus, $\xi=O_p(n)$. If $k \leq K-K_0$, since $\widehat{\pi}_k =O_p(n^{-1/2}/\log n)$, $\lambda=a/\sqrt{n}$ and $\epsilon=o(n^{-1/2}/\log n)$, we have
\begin{eqnarray*}
\left\{n\lambda\frac{1}{\epsilon+\widehat{\pi}_k}\right\} / n = \lambda \frac{1}{\epsilon+\widehat{\pi}_k} = O_p(\lambda\cdot n^{1/2}\log n) \rightarrow \infty
\end{eqnarray*}
with probability tending to one. Hence, the second term in (\ref{eq2}) dominates the first and third terms when $k \leq K-K_0$ and $\widehat{\pi}_k < n^{-1/2}/\log n$, which implies that (\ref{eq1}) holds or equivalently, $\widehat{\pi}_k = 0$, $k=1,\ldots,K-K_0$ with probability tending to one. This completes the proof of Theorem 2.

\subsection*{B. Tables and Graphs}

\begin{figure}
\centering
\subfigure[] {\includegraphics[width=2.9in]{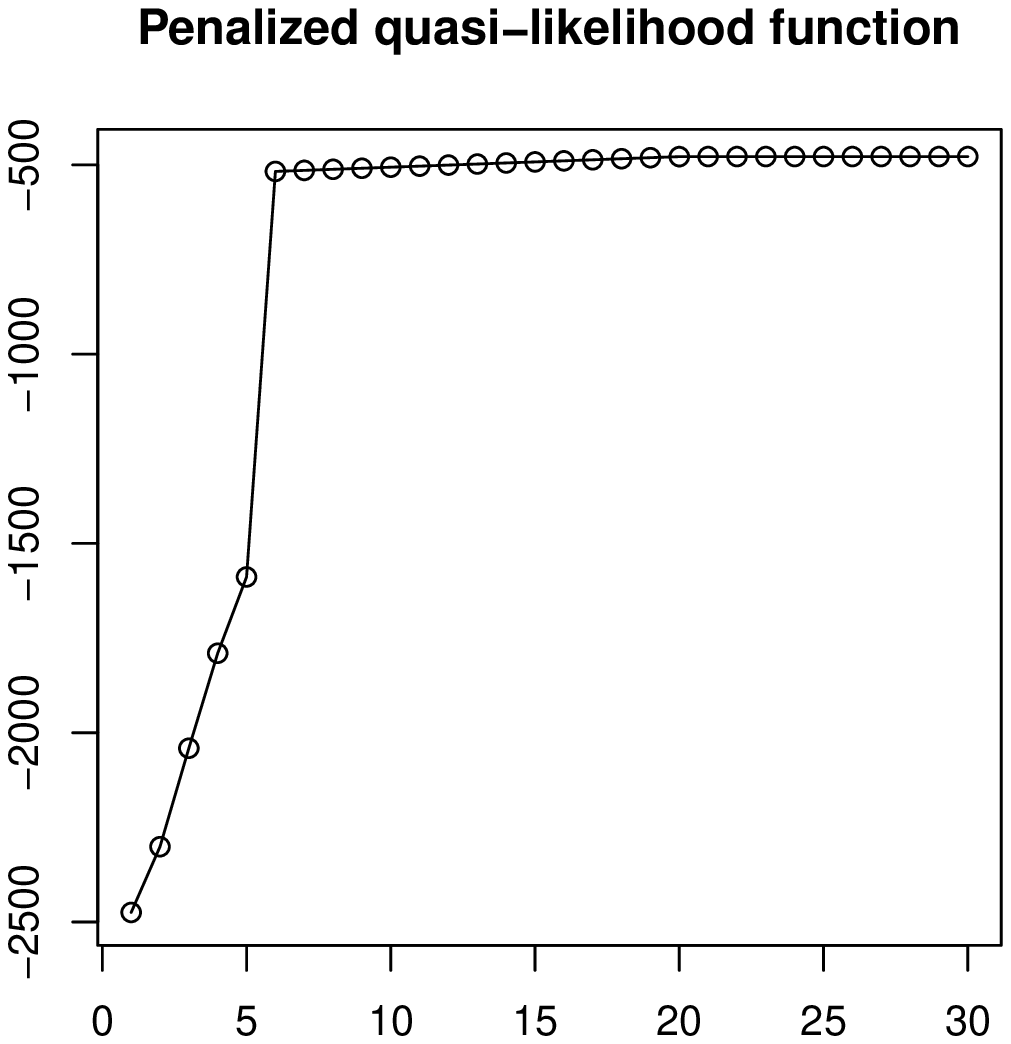}}\quad
\subfigure[]
{\includegraphics[width=3.1in]{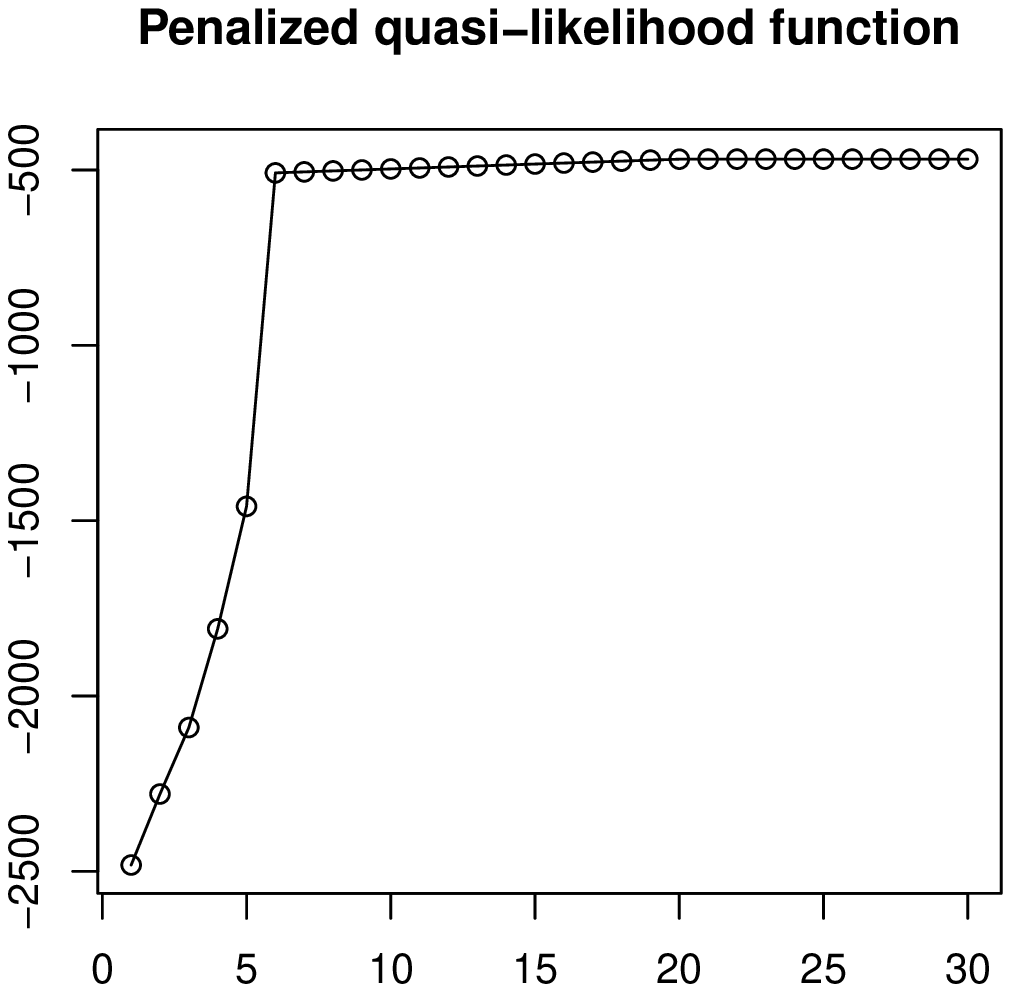}}
\caption{Evolutions of the penalized quasi-likelihood function for the simulated data set in Example 2 in one typical run: (a) $\rho=0.3$, (b) $\rho=0.6$.}
\label{figEx2}
\end{figure}

\begin{center} {
\begin{table}[htb!]
\caption{\label{tab4} {Estimation results in Example 2: (a) true values of mixture proportions and mixture parameters; (b) means of the parameter estimates; (c) means of the biases for the mixture proportions and mixture parameters; (d) mean squared errors (MSE) for the mixture proportions and mixture parameters. The values of bias and MSE are times 100. For the proposed PQL and PQL2 methods, the results below are summarized based on the models with correctly specified $K_0$ in 1000 replications.}
\hspace{2.5cm}}{{\normalsize \tabcolsep 0.075cm
\renewcommand{\arraystretch}{1.0}
\begin{tabular}{ccccccccccccc}
\hline\hline
Setting &  $\beta_{10}$ & $\beta_{11}$ & $\beta_{12}$ & $\beta_{13}$ & $\beta_{20}$ & $\beta_{21}$ & $\beta_{22}$ & $\beta_{23}$ & $\phi_1$ & $\phi_2$ & $\pi_1$ & $\pi_2$\\
\hline
$K_0=2$ & \multicolumn{12}{c}{True values}\\
 & 0 & 3 & -1 & 1 & 4 & -2 & 0 & 1 & 2 & 1 & 0.667 & 0.333\\
$\rho=0.3$ & \multicolumn{12}{c}{Mean}\\
PQL & 0.005 & 2.996 & -1.005 & 0.995 & 4.001 & -1.998 & -0.001 & 0.998 & 1.983 & 0.972 & 0.672 & 0.328\\
PQL2 & 0.005 & 2.997 & -1.006 & 0.995 & 4.001 & -1.998 & -0.001 & 0.998 & 2.000 & 0.989 & 0.672 & 0.328\\
QIFC & -0.013 & 3.014 & -1.010 & 1.000 & 4.001 & -2.000 & -0.001 & 0.999 & 2.005 & 0.977 & -- & --\\
 & \multicolumn{12}{c}{Bias}\\
PQL & 0.584 & -0.420 & -0.531 & -0.483 & 0.130 & 0.190 & -0.329 & -0.319 & -1.726 & -2.780 & 0.561 & -0.561\\
PQL2 & 0.502 & -0.296 & -0.555 & -0.456 & 0.149 & 0.180 & -0.320 & 0.315 & 0.015 & -1.085 & 0.561 & -0.561\\
QIFC & -1.282 & 1.405 & -0.997 & 0.033 & 0.117 & 0.006 & -0.222 & 0.417 & 0.458 & -2.254 & -- & --\\
 & \multicolumn{12}{c}{MSE}\\
PQL & 1.138 & 1.217 & 0.692 & 0.755 & 0.112 & 0.154 & 0.132 & 0.127 & 2.436 & 0.926 & 0.011 & 0.011\\
PQL2 & 1.029 & 1.082 & 0.603 & 0.703 & 0.101 & 0.135 & 0.117 & 0.112 & 2.531 & 0.892 & 0.011 & 0.011\\
QIFC & 1.152 & 1.177 & 0.654 & 0.749 & 0.112 & 0.151 & 0.131 & 0.125 & 2.748 & 0.912 & -- & --\\
$\rho=0.6$ & \multicolumn{12}{c}{Mean}\\
PQL & -0.002 & 2.997 & -1.002 & 1.003 & 4.002 & -2.000 & -0.001 & 0.997 & 1.980 & 0.981 & 0.680 & 0.320\\
PQL2 & -0.001 & 2.999 & -1.001 & 0.998 & 4.002 & -2.000 & -0.001 & 0.998 & 2.003 & 1.003 & 0.680 & 0.320\\
QIFC & -0.023 & 3.020 & -1.007 & 1.004 & 4.001 & -2.002 & -0.001 & 0.999 & 2.016 & 0.993 & -- & --\\
 & \multicolumn{12}{c}{Bias}\\
PQL & -0.204 & -0.298 & -0.175 & 0.296 & 0.215 & -0.033 & -0.085 & -0.289 & -1.951 & -1.878 & 1.334 & -1.334\\
PQL2 & -0.056 & -0.131 & -0.125 & -0.155 & 0.213 & -0.040 & -0.088 & -0.195 & 0.267 & 0.311 & 1.334 & -1.334\\
QIFC & -2.346 &  2.003 & -0.698 & 0.438 & 0.103 & -0.244 & -0.110 & -0.102 & 1.557 & -0.723 &  -- & --\\
 & \multicolumn{12}{c}{MSE}\\
PQL & 1.275 & 1.174 & 0.729 & 0.774 & 0.118 & 0.149 & 0.121 & 0.140 & 3.763 & 1.386 & 0.031 & 0.031\\
PQL2 & 0.910 & 0.751 & 0.468 & 0.438 & 0.072 & 0.089 & 0.061 & 0.077 & 4.175 & 1.428 & 0.031 & 0.031\\
QIFC & 1.101 & 0.907 & 0.518 & 0.491 & 0.083 & 0.100 & 0.071 & 0.089 & 4.489 & 1.410 & -- & --\\
\hline
\end{tabular}
}}
\end{table}
}
\end{center}

\begin{figure}
\centering
{\includegraphics[width=3.5in]{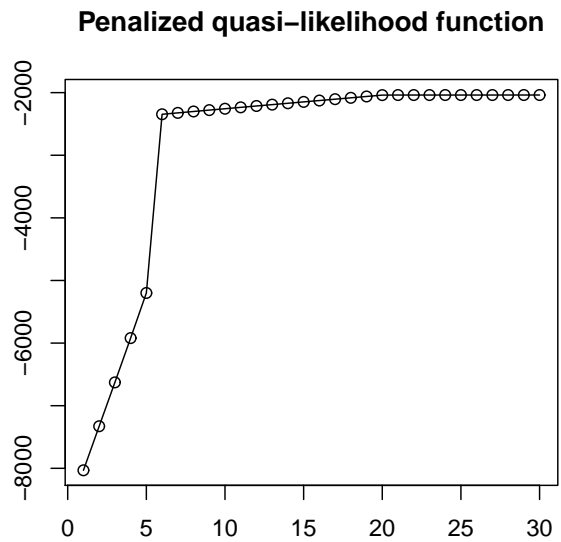}}
\caption{The Evolution of the penalized quasi-likelihood function for the simulated data set in Example 3 in one typical run.}
\label{figEx3}
\end{figure}

\begin{center}{
\begin{table}[htb!]
\caption{\label{tab5} {Estimation results in Example 3: (a) true values of mixture proportions and mixture parameters; (b) means of the parameter estimates; (c) means of the biases for the mixture proportions and mixture parameters; (d) mean squared errors (MSE) for the mixture proportions and mixture parameters. The values of bias and MSE are times 100. For the proposed PQL and PQL2 methods, the results above are summarized based on the models with correctly specified $K_0$ in 1000 replications.}
\hspace{5.5cm}}{{\normalsize \tabcolsep 0.27cm
\renewcommand{\arraystretch}{0.9}
\begin{tabular}{ccccccccccc}
\hline\hline
 & True & \multicolumn{3}{c}{PQL} & \multicolumn{3}{c}{PQL2} & \multicolumn{3}{c}{QIFC} \\
 & values & Mean & Bias & MSE & Mean & Bias & MSE & Mean & Bias & MSE\\
\hline
  $\beta_{10}$ & 2 & 1.992 & -0.752 & 0.493 & 1.993 & -0.698 & 0.368 & 1.996 & -0.360 & 0.341\\
 $\beta_{11}$ & 1 & 0.998 & -0.194 & 0.603 & 1.001 & -0.014 & 0.327 & 1.000 & 0.107 & 0.357\\
 $\beta_{12}$ & -1 &  -0.991 & 0.944 & 0.804 & -0.989 & 1.113 & 0.439 & -0.996 & 0.415 & 0.481\\
 $\beta_{13}$ & 1.5 & 1.496 & -0.392 & 0.626 & 1.496 & -0.436 & 0.296 & 1.499 & -0.213 & 0.330\\
 $\beta_{14}$ & 1 & 0.998 & -0.359 & 0.348 & 0.998 & -0.234 & 0.179 & 0.999 & -0.082  & 0.191\\
 $\beta_{20}$ & -4 & -3.988 & 1.219 & 0.277 & -3.991 & 0.693 & 0.211 & -3.999 & 0.071 & 0.215\\
 $\beta_{21}$ & 2 & 1.994 & -0.621 & 0.388 & 1.995 & -0.453 & 0.215 & 1.998 & -0.350 & 0.227\\
 $\beta_{22}$ & 1 & 1.001 & 0.294  & 0.535 & 1.002 & 0.162  & 0.291 & 1.001 & 0.246 & 0.312\\
 $\beta_{23}$ & -2 & -2.001 & -0.098 & 0.397 & -2.000 & 0.019 & 0.211 & -2.001 & -0.112 & 0.226\\
 $\beta_{24}$ & 0 & -0.002 & -0.188 & 0.213 & 0.001 & 0.069 & 0.111 & 0.001 & 0.121 & 0.119\\
 $\beta_{30}$ & -2 & -1.998 & 0.249 & 0.140 & -1.997 & 0.268 & 0.123 & -1.999 & 0.235 & 0.135\\
 $\beta_{31}$ & -2 & -1.999 & 0.084 & 0.210 & -2.000 & 0.046 & 0.169 & -1.999 & 0.088 & 0.185\\
 $\beta_{32}$ & 1 & 0.999 & -0.064 & 0.292 & 1.000 & -0.009 & 0.245 & 1.001 & -0.060 & 0.262\\
 $\beta_{33}$ & 0 & -0.001 & -0.065 & 0.199 & 0.000 & -0.039 & 0.163 & 0.000 & -0.081 & 0.181\\
 $\beta_{34}$ & 1 & 1.000 & -0.090 & 0.106 & 1.000 & -0.012 & 0.082 & 1.000 & -0.008  & 0.090\\
 $\beta_{40}$ & 0 & -0.006 & -0.629 & 0.219 & -0.008 & -0.776 & 0.197 & -0.001 & -0.082 & 0.213\\
 $\beta_{41}$ & 1 & 0.998 & -0.193 & 0.320 & 0.998 & -0.194 & 0.260 & 0.999 & -0.101 & 0.269\\
 $\beta_{42}$ & 0 & -0.010 & -0.988 & 0.417 & -0.005 & -0.496 & 0.340 & 0.001 & 0.070 & 0.358\\
 $\beta_{43}$ & 1 & 1.006 & 0.582 & 0.295 & 1.003 & 0.199 & 0.254 & 1.001 & 0.081 & 0.270\\
 $\beta_{44}$ & 1 & 1.001 & 0.140 & 0.163 & 1.002 & 0.174 & 0.135 & 1.001 & 0.109 & 0.144\\
 $\beta_{50}$ & -4 & -3.998 & 0.200 & 0.465 & -4.001 & 0.001 & 0.463 & -4.000 & 0.042 & 0.469\\
 $\beta_{51}$ & 0 & -0.001 & -0.050 & 0.883 & 0.000 & -0.044 & 0.883 & -0.005 & -0.490 & 0.865\\
 $\beta_{52}$ & -1 & -0.992 & 0.806 & 1.319 & -0.992 & 0.775 & 1.321 & -0.995 & 0.882 & 1.265\\
 $\beta_{53}$ & -1 & -1.005 & -0.416 & 0.958 & -1.005 & -0.338 & 0.965 & -1.003 & -0.338 & 0.936\\
 $\beta_{54}$ & -1.5 & -1.483 & 0.654 & 0.451 & -1.493 & 0.651  & 0.454 & -1.497 & 0.348 & 0.443\\
 $\sigma^2_1$ & 0.5 & 0.494 & -0.592 & 0.189 & 0.493 & -0.699 & 0.189 & 0.495 & -0.578 & 0.150\\
 $\sigma^2_2$ & 0.3 & 0.292 & -0.839 & 0.059 & 0.292 & -0.824 & 0.059 & 0.297 & -0.294 & 0.052\\
 $\sigma^2_3$ & 0.1 & 0.097 & -0.251 & 0.008 & 0.098 & -0.237 & 0.008 & 0.099 & -0.137 & 0.008\\
 $\sigma^2_4$ & 0.15 & 0.143 & -0.873 & 0.023 & 0.143 & -0.735 & 0.023 & 0.147 & -0.354 & 0.017\\
 $\sigma^2_5$ & 0.6 & 0.604 & 0.383 & 0.168 & 0.601 & 0.144 & 0.165 & 0.596 & -0.396 & 0.146\\
 $\pi_1$ & 0.25 & 0.268 & 1.797 & 0.054 & 0.268 & 1.797 & 0.054 & -- & -- & --\\
 $\pi_2$ & 0.25 & 0.264 & 1.418 & 0.037 & 0.264 & 1.418 & 0.037 & -- & -- & --\\
 $\pi_3$ & 0.15 & 0.132 & -1.791 & 0.050 & 0.132 & -1.791 & 0.050 & -- & -- & --\\
 $\pi_4$ & 0.15 & 0.132 & -1.796 & 0.054 & 0.132 & -1.796 & 0.054 & -- & -- & --\\
 $\pi_5$ & 0.2 & 0.204 & 0.373 & 0.003 & 0.204 & 0.373 & 0.003 & -- & -- & --\\
 \hline
\end{tabular}
}}
\end{table}
}
\end{center}

\begin{figure}
\centering
{\includegraphics[width=3.5in]{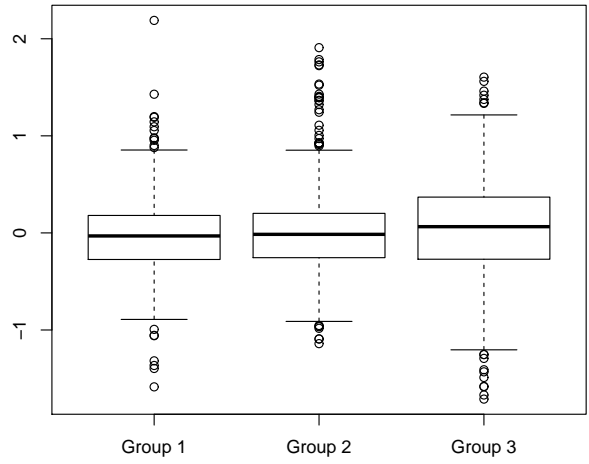}}
\caption{The boxplots of residulas for lbili under the fitted LMMs.}
\label{fig-LMM}
\end{figure}

\begin{figure}
\centering
{\includegraphics[width=3.5in]{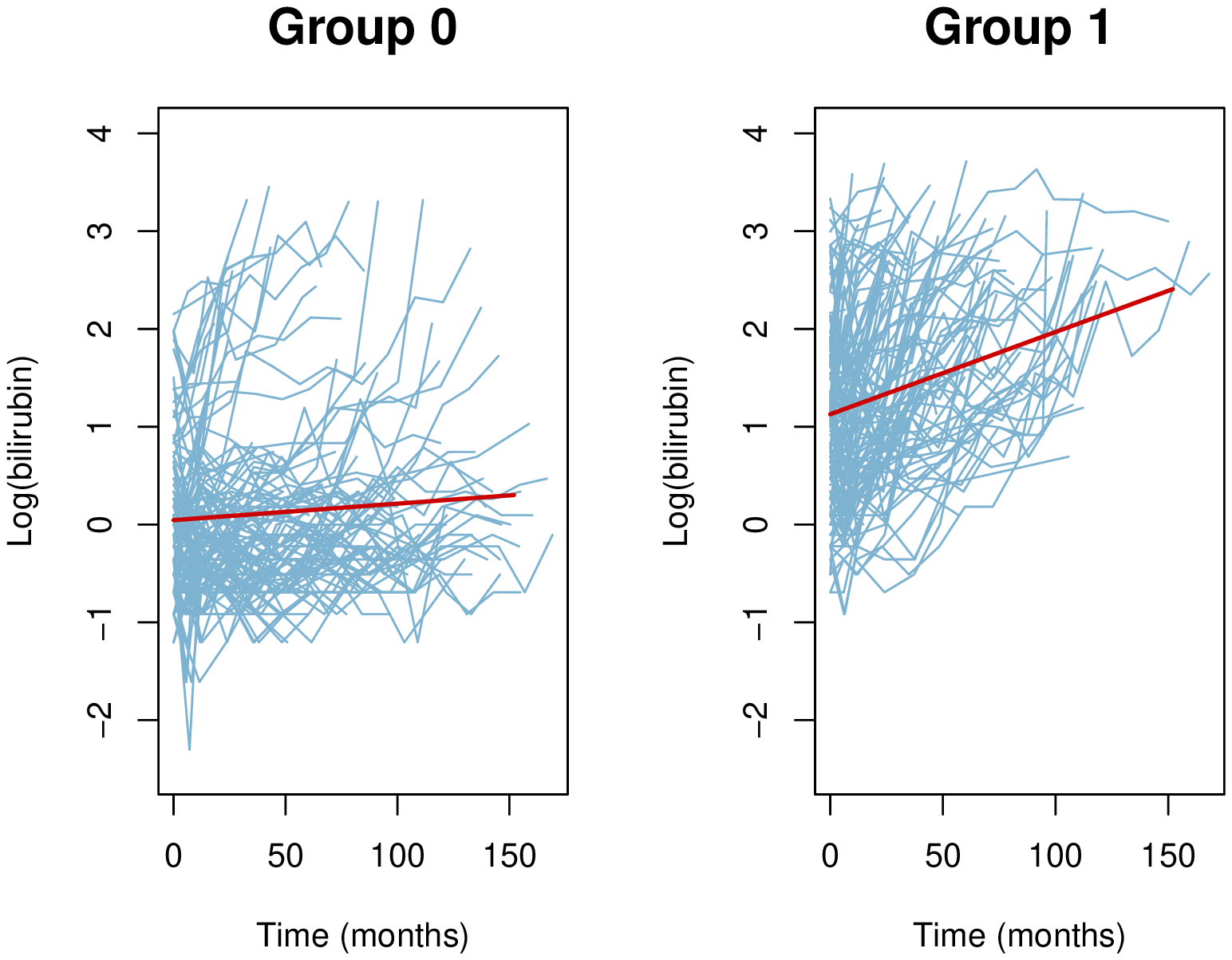}}
\caption{Trajectories plots for the PBC data. Observed evolution of lbili marker for 312 patients. The red lines show the fitted mean profiles in two groups.}
\label{fig-PQL}
\end{figure}


\begin{thebibliography}{}

\bibitem{}
Bollerslev, T. and Wooldridge, J.M. (1992). Quasi-maximum likelihood estimation and inference in dynamic models with time-varying covariances. \textit{Econometric Reviews} \textbf{11,} 143--172.

\bibitem{}
Booth, J.G., Casella, G., and Hobert, J.P. (2008). Clustering using objective functions and stochastic search. \textit{Journal of the Royal Statistical Society} \textbf{B70,} 119--139.


\bibitem{}
Celeux, G., Martin, O., and Lavergne, C. (2005). Mixture of linear mixed models for clustering gene expression profiles from repeated microarray experiments. \textit{Statistical Modelling} \textbf{5,} 243--267.

\bibitem{}
Chen, J. and Khalili, A. (2008). Order selection in finite mixture models with a nonsmooth penalty. \textit{Journal of the American Statistical Association} \textbf{104,} 187--196.

\bibitem{}
Dacunha-Castelle, D. and Gassiat, K. (1997). Testing in locally conic models and application to mixture models.  \textit{ESAIM: Probability and Statistics} \textbf{1,} 285--317.


\bibitem{}
Dacunha-Castelle, D. and Gassiat, K. (1999). Testing the order of a model using locally conic parametrization: population mixtures and stationary ARMA processes. \textit{The Annals of Statistics} \textbf{27,} 1178--1209.


\bibitem{}
Dasgupta, A. and Raftery, A.E. (1998). Detecting features in spatial point processes with clutter via model-based clustering. \textit{Journal of the American Statistical Association} \textbf{93,} 294--302.

\bibitem{}
De la Cruz-Mes\'{\i}a, R., Quintana, F.A., and Marshall, G. (2008). Model-based clustering for longitudinal data. \textit{Computational Statistics $\&$ Data Analysis} \textbf{52,} 1441--1457.

\bibitem{}
Dickson, E.R., Grambsch, P.M., Fleming, T.R., Fisher, L.D., and Langworthy, A. (1989). Prognosis in primary biliary cirrhosis: Model for decision making. \textit{Hepatology} \textbf{10,} 1--7.


\bibitem{}
Erosheva, E.A., Matsueda, R.L., and Telesca, D. (2014). Breadking bad: two decades of life-course data analysis in criminology, developmental psychology, and beyond. \textit{Annual Review of Statistics and Its Application} \textbf{1,} 301--332.

\bibitem{}
Fan, J. and Li, R. (2001). Variable selection via nonconcave penalized likelihood and its oracle properties. \textit{Journal of the American Statistical Association} \textbf{96,} 1348--1360.

\bibitem{}
Ferguson, T.S. (1996). A course in large sample theory. Chapman $\&$ Hall.

\bibitem{}
Fraley, C. and Raftery, A.E. (2002). Model-based clustering discriminant analysis and density estimation. \textit{Journal of the American Statistical Association} \textbf{97,} 611--631.

\bibitem{}
Genolini, C. and Falissard, B. (2010). KmL: k-means for longitudinal data. \textit{Computational Statistics} \textbf{25,} 317--328.


\bibitem{}
Heinzl, F. and Tutz, G. (2013). Clustering in linear mixed models with approximate Dirichlet process mixtures using EM algorithm. \textit{Statistical Modelling} \textbf{13,} 41--67.

\bibitem{Heinzl(2014)}
Heinzl, F. and Tutz, G. (2014). Clustering in linear-mixed models with a group fused lasso penalty. \textit{Biometrical Journal} \textbf{56,} 44--68.

\bibitem{}
Hennig, C. (2004). Breakdown points for maximum likelihood estimators of location-scale mixtures. \textit{The Annals of Statistics} \textbf{32,} 1313--1340.

\bibitem{}
Hoofnagle, J.H., David, G.L., Schafer, D.F., Peters, M., Avigan, M.I., Pappas, S.C., Hanson, R.G., Minuk G.Y., Dusheiko, G.M., and Campbell, G. (1986). Randomized trial of chlorambucil for primary biliary cirrhosis. \textit{Gastroenterology} \textbf{91,} 1327-1334.

\bibitem{}
Huang, J.Z., Zhang, L., and Zhou, L. (2007). Efficient estimation in marginal partially linear models for longitudinal/clustered data using splines. \textit{Scandinavian Journal of Statistics} \textbf{34,} 451--477.

\bibitem{}
Huang, T., Peng, H., and Zhang, K. (2016). Model selection for Gaussian mixture models. \textit{Statistica Sinica}, in press.

\bibitem{}
Keribin, C. (2000). Consistent estimation of the order of mixture models. \textit{Sankhy$\bar{a}$} \textbf{62,} 49--66.

\bibitem{}
Kom\'{a}rek, A. and Kom\'{a}rkov\'{a}, L. (2013). Clustering for multivariate continuous and discrete longitudinal data. \textit{The Annals of Applied Statistics} \textbf{7,} 177--200.

\bibitem{}
Kom\'{a}rek, A. and Lesaffre, E. (2008). Generalized linear mixed model with a penalized Gaussian mixture as a random effects distribution. \textit{Computational Statistics $\&$ Data Analysis} \textbf{52,} 3441--3458.

\bibitem{}
Leroux, B. (1992). Consistent estimation of a mixing distribution. \textit{The Annals of Statistics} \textbf{20,} 1350--1360.

\bibitem{}
Liang, K.Y. and Zeger, S.L. (1986). Longitudinal data analysis using generalised linear models. \textit{Biometrika} \textbf{73,} 12--22.


\bibitem{}
Maruotti, A. (2011). Mixed hidden Markov models for longitudinal data: an overview. \textit{International Statistical Review} \textbf{79,} 427--454.

\bibitem{}
McNicholas, P.D. and Murphy, T.B. (2010). Model-based clustering of longitudinal data. \textit{The Canadian Journal of Statistics} \textbf{38,} 153--168.

\bibitem{}
Murtaugh, P.A., Dickson, E.R., van Dam, G.M., Malinchoc, M., Grambsch, P.M., Langworthy, A.L., and Gips, C.H. (1994). Primary biliary cirrhosis: prediction of short-term survivial based on repeated patient visits. \textit{Hepatology} \textbf{20,} 126--134.

\bibitem{}
Pickles, A. and Croudace, T. (2010). Latent mixture models for multivariate and longitudinal outcomes. \textit{Statistical Methods in Medical Reseaerch} \textbf{19,} 271--289.

\bibitem{}
Pontecorvo, M.J., Levinson J.D., and Roth, J.A. (1992). A patient with primary biliary cirrhosis and multiple sclerosis. \textit{The American Journal of Medicine} \textbf{92,} 433--436.

\bibitem{}
Roeder, K. and Wasserman, L. (1997). Practical density estimation using mixtures of normals. \textit{Journal of the American Statistical Association} \textbf{92,} 894--902.

\bibitem{}
Schwarz, G. (1978). Estimating the dimension of a model. \textit{Annals of Statistics} \textbf{6,} 461--464.

\bibitem{}
Talwalkar, J.A. and Lindor, K.D. (2003). Primary biliary cirrhosis. \textit{Lancet} \textbf{362,} 53--61.

\bibitem{}
Verbeke, G. and Lesaffre, E. (1996). A linear mixed-effects model with heterogeneity in the random-effects population. \textit{Journal of the American Statistical Association} \textbf{91,} 217--221.

\bibitem{}
Wang, H., Li, R., and Tsai, C.L. (2007). Tuning parameter selelctors for the smoothly clipped absolute deviation method. \textit{Biometrika} \textbf{94,} 553--568.

\bibitem{}
Wang, L. (2011). GEE analysis of clustered bianry data with diverging number of covariates. \textit{The Annals of Statistics} \textbf{39,} 389--417.

\bibitem{}
Wang, X. and Qu, A. (2014). Efficient classification for longitudinal data. \textit{Computatioinal Statistics $\&$ Data Analysis} \textbf{78,} 119--134.

\bibitem{}
Xu, P., Zhang, J., Huang, X., and Wang, T. (2016). Efficient estimation of marginal generalized partially linear single-index models with longitudinal data. \textit{TEST} \textbf{25,} 413--431.

\bibitem{}
Xu, P. and Zhu, L. (2012). Estimation for a marginal generalized single-index longitudinal model. \textit{Journal of Multivariate Analysis} \textbf{105,} 285--299.

\bibitem{}
Yao, W. (2015). Label switching and its solutions for frequentist mixture model. \textit{Journal of Statistical Computation and Simulation} \textbf{85,} 1000--1012.

\bibitem{}
Zhu, W. and Fan, Y. (2016). Relabelling algorithms for mixture models with applications for large data sets.  \textit{Journal of Statistical Computation and Simulation} \textbf{86,} 394--413.

\end{thebibliography}
\end{document}